\documentclass{aastex} 
\usepackage{emulateapj5} 

\newcommand{\n}{\nodata}
\newcommand{\rotm}{$\ \rm rad \ m^{-2}$}

\def\gtrsim{\mathrel{\hbox{\rlap{\hbox{\lower4pt\hbox{$\sim$}}}\hbox{$>$}}}}
\def\lesssim{\mathrel{\hbox{\rlap{\hbox{\lower4pt\hbox{$\sim$}}}\hbox{$<$}}}}

\slugcomment{Accepted for publication in the {\it Astrophysical Journal}.}
\shortauthors{Lister \& Smith}
\shorttitle{High- and Low-Optically Polarized Radio Quasars}

\begin{document}
\title{Intrinsic Differences in the Inner Jets of \\ High-
and Low-Optically Polarized Radio Quasars}
\author{Matthew L. Lister}
 
\affil{Jet Propulsion Laboratory, California Institute of Technology
\\ MS 238-332, 4800 Oak Grove Drive, Pasadena, CA 91109-8099
\\ \email{lister@hyaa.jpl.nasa.gov}}

\and

\author{Paul S. Smith}
\affil{National Optical Astronomy Observatories
\\ P.O. Box 26732, 950 North Cherry Avenue,  Tucson, AZ 85726-6732
\\ \email{psmith@noao.edu}}

\begin{abstract}
A significant fraction of flat-spectrum, radio-loud quasars display
most of the characteristics of relativistically beamed, high-optical
polarization blazars, yet are weakly polarized in the optical regime
($m_{\rm opt} \lesssim 3\%$). We have conducted a high-resolution
polarization study with the VLBA at 22 and 43 GHz to look for
differences in the parsec-scale magnetic field structures of 18 high-
and low-optically polarized, compact radio-loud quasars (HPQs and
LPRQs, respectively). We find a strong correlation between the
polarization level of the unresolved parsec-scale radio core at 43 GHz
and overall optical polarization of the source, which suggests a
common (possibly co-spatial) origin for the emission at these two
wavelengths. The electric vectors of the polarized 43 GHz radio cores
are roughly aligned with the inner jet direction, indicating magnetic
fields perpendicular to the flow. Similar orientations are seen in the
optical, suggesting that the polarized flux at both wavelengths is due
to one or more strong transverse shocks located very close to the base
of the jet. In LPRQs, these shocks appear to be weak near the core,
and gradually increase in strength down the jet. The LPRQs in our
sample tend to have less luminous radio cores than the HPQs, and jet
components with magnetic fields predominantly parallel to the jet. The
components in HPQ jets, on the other hand, tend to have perpendicular
magnetic field orientations. These differences cannot be accounted for
by a simple model in which HPQs and LPRQs are the same type of object,
seen at different angles to the line of sight.  A more likely scenario
is that LPRQs represent a quiescent phase of blazar activity, in which
the inner jet flow does not contain strong shocks. Our high-resolution
observations have shown that high rotation measures (up to 3000 $\rm
rad\ m^{-2}$) previously seen in the nuclear regions of HPQs are
present in LPRQs as well.  The low-redshift quasars in our sample tend
to have jet components with larger 43/22 GHz depolarization ratios
than those found in the high-redshift sources.  This may be due to
small-scale magnetic field fluctuations in the Faraday screens that
are being smeared out in the high-redshift sources by the poorer
spatial resolution of the restoring beam.

\end{abstract}

\keywords{
galaxies : active --- 
galaxies : jets --- 
galaxies: magnetic fields ---
quasars : general ---
polarization ---
shock waves}
 
\section{Introduction}

Since its introduction in 1978, the term ``blazar'' has been
synonymous with radio-loud active galactic nuclei that have steep,
smooth optical continua, highly core-dominated radio morphologies, and
fluxes that are highly variable at all wavelengths. The bulk of their
radiation is thought to be highly relativistically beamed synchrotron
emission from plasma outflows in the form of jets.

An important defining characteristic of blazars is their high degree
of polarization at radio through optical wavelengths. This
polarization is quite variable, often on short time-scales ($\lesssim
1$ day), and indicates spatially small emitting regions with
well-ordered magnetic fields. These are likely formed by relativistic
shock fronts which re-order an originally tangled magnetic field in
the jet. This shock model (e.g., \citealt*{HAA85};
\citealt{MG85}) has successfully reproduced many of the observed
radio properties of blazar jets.

One of the best indicators of blazar activity is the
level of fractional polarization in the optical regime
($m_{\rm opt}$). Studies of large AGN samples (e.g.,
\citealt*{SMA84}; \citealt{BSW90}) have shown that nearly all ``normal''
(i.e., radio-quiet) quasars have very weak, non-variable optical
polarizations ($\sim 0.5 \%$). With the exception of a few sources
such as OI 287, whose high polarizations can be attributed to
scattering \citep[e.g.,][]{RS88}, the optical polarizations of
radio-quiet quasars rarely exceed $\sim 3 \%$.  Those of blazars, on
the other hand, span a large range, up to $\sim 46$\% in some cases
\citep{MBB90}, and are attributed to synchrotron emission
from their relativistic jets. This dichotomy has led to the
classification scheme ``high-optically polarized quasar'' (HPQ) for
sources with $m_{\rm opt} > 3\%$ (i.e., blazars) and ``low-optically
polarized quasar'' (LPQ) for those AGNs with $m_{\rm opt} < 3\%$.

The connection between high-optical polarization and jet synchrotron
emission might suggest that all core-dominated, radio-loud AGNs
containing relativistically beamed jets should be HPQs, but this is
not the case. Many blazars have optical polarizations that
occasionally dip below $3\%$, and there are many other radio-loud AGNs
that display most of the characteristics of blazars, but have
consistently low optical polarizations.  Perhaps the most famous
example of a low-optically polarized, radio-loud quasar (LPRQ) is that
of 3C~273, a well-studied superluminal source, whose optical
polarization has rarely exceeded 3\%. High-sensitivity
photo-polarimetry of this object by \citet{IMT89} revealed a
``mini-blazar'' component, whose overall contribution to the optical
flux is swamped by a strong optical continuum, possibly from a large,
hot accretion disk.  If its blazar component were significantly
stronger, 3C~273 would in all likelihood have the properties of a
typical HPQ.

The reason why all radio-loud AGNs are not HPQs may lie with their
relativistic jets, as there is ample evidence showing a link between
the optical polarization and radio jet properties of blazars. For
example, the optical electric polarization vectors of radio-loud AGNs
are known to be well-aligned with their parsec-scale
\citep{RS85} and kiloparsec-scale
\citep*{SAM79} jets, indicating a shared radio and optical (and
possibly co-spatial) emission mechanism. This view has been supported
by variability studies such as that of \citet{HB92} and \citet{VVS91},
who found correlated flaring activity at optical and radio
wavelengths. Also, \citet*{GSS96} studied a small sample of blazars
and found weak evidence that the levels of optical and parsec-scale
radio polarization were correlated.

Some authors \citep[e.g.,][]{F88} have speculated that LPRQs and HPQs
may represent the quiescent and active phases, respectively, of the
same object. Others have suggested that orientation plays an important
role, with the jets of LPRQs being oriented farther from the line of
sight \citep{VTU92}. The latter model is supported somewhat by
observations showing that LPRQs are generally less variable
(\citealt{VTU92}), and have smaller misalignments between their
parsec- and kiloparsec-scale jet directions \citep{ILT91,XU94}.

In this paper, we present observations that show a direct link between
the optical polarization and parsec-scale radio properties of compact,
radio-loud AGNs. We also show that there are intrinsic differences in
the jets of LPRQs and HPQs that cannot be explained purely by
differences in orientation.  These intrinsic differences are
associated with the magnetic field structure of the parsec-scale jet,
which are in turn responsible for the optical-through-radio
polarization properties of compact radio quasars.

\section{Sample selection}

In order to investigate potential parsec-scale differences in LPRQs
and HPQs, we assembled a sample of nine known LPRQs that have
sufficient compact radio flux to be imaged by the NRAO\footnote{The
National Radio Astronomy Observatory is a facility of the National
Science Foundation, operated under cooperative agreement by Associated
Universities Inc.} Very Long Baseline Array (VLBA) in snapshot mode at
22 and 43 GHz. The relatively low source opacity and high spatial
resolution at these observing frequencies allow us to probe regions
very close to the base of the jet, where the electron energies are
likely to be high enough to produce large amounts of optical
synchrotron emission.

Candidate objects were drawn from a complete list of northern
optically-identified quasars \citep*{SMK94} with declination $> 0^o$,
5 GHz flux density $> 1$ Jy, and $\alpha > -0.5$, where $\alpha$ is
the spectral index between 1.4 and 5 GHz ($S \sim \nu^\alpha$).  We
reduced the chance of including ``mis-identified'' LPRQs that may in
fact be high-polarization objects by restricting our final sample to
ten objects whose optical polarization has never exceeded 3\% at three
or more epochs. Our subsequent optical polarization measurements (see
\S\ref{optical_obs}) revealed that one of these objects (1633+382)
should now be re-classified as an HPQ, leaving a total of nine LPRQs in
our final sample.

We assembled a complementary sample of HPQs from those recently
observed at 43 GHz with the VLBA by \citet*{LMG98}, and
\citet{M99}. Adding our observations of 1633+382 and the calibrator
source 3C~279 gave us a matched sample of nine HPQs.

Although these final HPQ and LPRQ samples are not statistically
complete, they are well-suited for making cross-comparisons between
these two classes of object, as their overall radio properties are
very similar. There are no significant differences in their redshift,
total 22 and 43 GHz radio luminosity, spectral index, and optical
magnitude distributions, according to Kolmogorov-Smirnov tests.  We
list the general properties of our samples in Table~\ref{totaldata}.

Throughout this paper we use a standard Freidmann cosmology with
deceleration parameter $q_o = 0.1$, zero cosmological constant
($\Lambda$) and Hubble constant $h = 0.65$, in units of $100 \ \rm km\
s^{-1}\ Mpc^{-1}$. We define the spectral index such that flux density
$S_\nu$ is proportional to $\nu^{\alpha}$, and give all position
angles in degrees east of north.

\section{Observations and data reduction}
\subsection{Radio observations}

Our radio observations were carried out at 22 and 43 GHz using all ten
antennas of the VLBA on UT 1999 January 12-14.  The
data were recorded in eight baseband channels (IFs) using 1-bit
sampling, with each IF having a bandwidth of 8 MHz.  Both right and
left hand polarizations were recorded simultaneously in IF pairs,
giving a total observing bandwidth of 32 MHz. Due to a receiver
failure, no 22 GHz data were gathered with the antenna at St. Croix.

The data were correlated using the VLBA correlator in Socorro, NM, and
subsequent data editing and calibration were performed at JPL using the
Astronomical Image Processing System (AIPS) software supplied by
NRAO. The calibration procedure followed that of the AIPS Cookbook
\citep{N90} and \citet*{LZD95}.

We calibrated our visibility amplitudes using the system temperatures
measured at each antenna, along with gain curves supplied by the
NRAO. We performed an opacity correction using single dish flux data
for several program sources, obtained concurrently at the Mets\"ahovi
Radio Observatory and kindly provided to us by H. Ter\"asranta. The
Mets\"ahovi fluxes for 2145+067 at 37 GHz and 4C~39.25 at 22 GHz were
used to establish the absolute flux density scale of our data at 22
and 43 GHz. These sources are both sufficiently compact (angular sizes
$< 4\arcsec$ at 1.4 GHz; \citealt*{MBP93}) that very little flux
should be resolved out in our VLBA images. We estimate our absolute
flux density scaling to be accurate to within $\sim 20\%$.

We determined the polarization leakage factors (also called
``D-factors'') for each antenna at 22 and 43 GHz using the AIPS task
LPCAL. This program uses a source model to distinguish between true
polarized signal and instrumental noise in the self-calibrated data,
and works best on nearly unpolarized sources, or those with simple
polarization structure. We ran this task on all of our program
sources, and found that at 43 GHz, the sources 1633+382, 4C~39.25, and
2145+067 provided the most self-consistent D-factor solutions, which
ranged up to $\sim 5$\%, with a scatter of $\sim 0.2 - 0.7$\%. At 22
GHz, we used NRAO~140 in lieu of 4C~39.25. For each observing
frequency, we averaged the antenna D-factor solutions obtained using
these sources to create a final set, and applied these to the data

The last step in the calibration was to determine the absolute R-L
phase difference at the reference antenna. This number is related to
the constant offset of the electric vector position angles ($\chi$) on
the sky from their true values. We used the known $\chi$'s of 4C~39.25,
as measured by \citet{A99} at 22 and 43 GHz on UT 1998 October
25 and UT 1999 February 8, to calibrate our $\chi$'s. The measurements of
\citet{A99} were originally calibrated using simultaneous polarization
measurements with the NRAO's Very Large Array (VLA). As an additional
check on our calibration, we found good agreement between our $\chi$'s
for 3C~279 and those of \citet{M99} at 43 GHz, who have been
monitoring the polarization of this source at two month intervals
throughout 1998 and 1999. Based on these comparisons, we estimate that
our absolute $\chi$ calibration is accurate to within $\sim 5\arcdeg$.

\subsection{Model fitting and analysis of radio data}

We used the Caltech DIFMAP package \citep{SPT94} to make Stokes I, Q, and U
images of our sources, which we then imported into AIPS to make our
final contour images, shown in Figures 1--11. A complete summary of our
image parameters is given in Tables~\ref{Kmapdata} and
\ref{Qmapdata}. We used a natural visibility weighting scheme for
those sources with more diffuse jet structure, as this gives better
sensitivity to weak emission.

We performed Gaussian model fits to each source in the uv plane at 22
and 43 GHz using the task ``modelfit'' in DIFMAP. The results of these
fits are given in Tables~\ref{coredata} and \ref{components}, and are
intended as a general guideline for interpreting the polarization
structure. Those components that we detected only at 43 GHz (due to
the higher angular resolution) are indicated with lowercase
letters. We caution that this type of model fitting does not always
produce unique results, especially for regions of nearly continuous
jet emission. Our procedure was to start with a model composed
entirely of CLEAN components, and gradually replace those in the inner
few milliarcseconds with Gaussian model components. We added components
farther down in the jet in regions where polarized flux was
present. While an accurate parameterization of errors is somewhat
difficult with this method, we estimate the given positions of strong,
isolated components to be accurate to within a quarter of a
beam width. For our measured flux densities, we estimate our errors to
be $\lesssim 10\%$, in accordance with \citet{GMA99} and
\citet{MJV99}. Our fits are less reliable for very weak components, and
those located in regions of diffuse emission.

In Table~\ref{coredata} we list the fitted radio core component
properties and optical data of our sample sources. We discuss the
latter (columns 6, 9, 12, and 15) in \S\ref{optical_obs}.  Columns 2
and 3 give the total core flux density in mJy at 22 and 43 GHz,
respectively. We tabulate the core luminosity in column 4, assuming a
spectral index $\alpha = 0$.  Column 5 gives the ratio of core to
extended (parsec-scale) flux density at 43 GHz, while columns 7 and 8
indicate the percentage polarization $m$ at the central pixel of the
core component at 22 and 43 GHz.  The percentage polarization is an
indicator of the degree of order in the magnetic field, and is defined
as $100 \times P/I$, where $P=({Q^2+U^2})^{1/2}$, and $Q$, $U$, and
$I$ are Stokes flux densities. We note that in some cases the peak of
polarized emission is offset from that of the $I$ emission, so that
the values of $m$ given in Tables~\ref{coredata} and
\ref{components} may not represent the maximum percentage polarization
associated with a particular component.  Since all of our polarization
detections in Tables~\ref{coredata} and
\ref{components} are $\gtrsim 5 \sigma$, we include no corrections for
Ricean bias \citep{WK74}, as these are all negligible.  In columns 10
and 11 we list the electric polarization vector position angle $\chi$
for those cores with detected polarized flux at 22 and 43 GHz.

In column 13, we list the ratio of percentage polarization at 43 GHz
to that at 22 GHz for those components detected at both
frequencies. In order to calculate these values, we matched the
uv-coverages as best as possible by flagging visibilities, and using
the same restoring beam in the 22 and 43 GHz images. For most sources,
it was necessary to shift the re-convolved images so the jet features
were coincident at both frequencies. This is due to the
self-calibration procedure, which moves the phase reference center to
the brightest feature in the image. At 43 GHz, many of the bright
cores were resolved into two or more components, which shifted the
apparent position of the brightest component by a small, but
non-negligible amount.  An exact determination of these shifts is
difficult, due to the large difference in observing frequencies, and a
general lack of steep-spectrum features along the jet with which to
register the images. The quantities most affected by registration
errors are the spectral index and fractional polarization ratio. The
errors in electric vector rotation are not as large, since this
quantity does not vary as much from pixel-to-pixel in our images.

We list the measured rotation in $\chi$ between 22 and 43 GHz for the
core components in column 14. These values were also obtained using
the convolved 43 GHz images. We have applied no overall rotation
measure corrections to the data in Tables~\ref{coredata} and
\ref{components}.

In Table~\ref{components}, we list the same data for the polarized jet
components in our sample sources. Columns 2 through 5 give the
position with respect to the core, and total flux density of each
component at 22 and 43 GHz. Steep-spectrum components that were too
faint to be detected at 43 GHz are indicated with a dagger.

\begin{figure*}
\epsscale{1}
\plotone{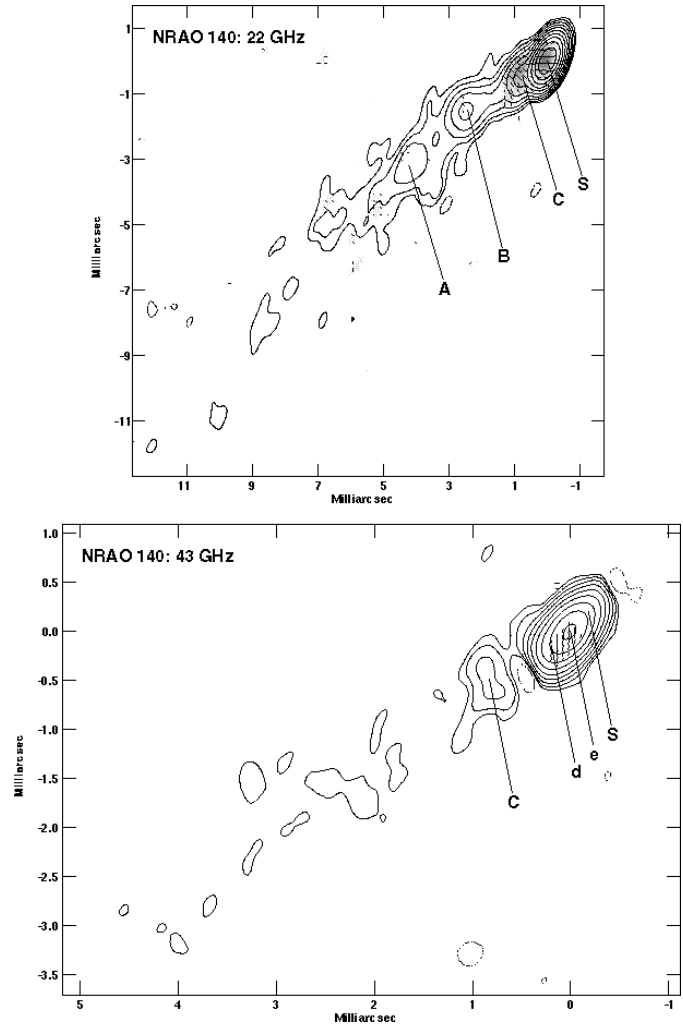}
\caption{VLBA total intensity images of NRAO 140 at 22
and 43 GHz, epoch 1999.03, with electric polarization vectors
superimposed. The greyscale in the 22 GHz image indicates linearly
polarized intensity ($({Q^2+U^2})^{1/2}$), where Q and U are the
Stokes flux densities. }
\end{figure*}

\begin{figure*}
\epsscale{1}
\plotone{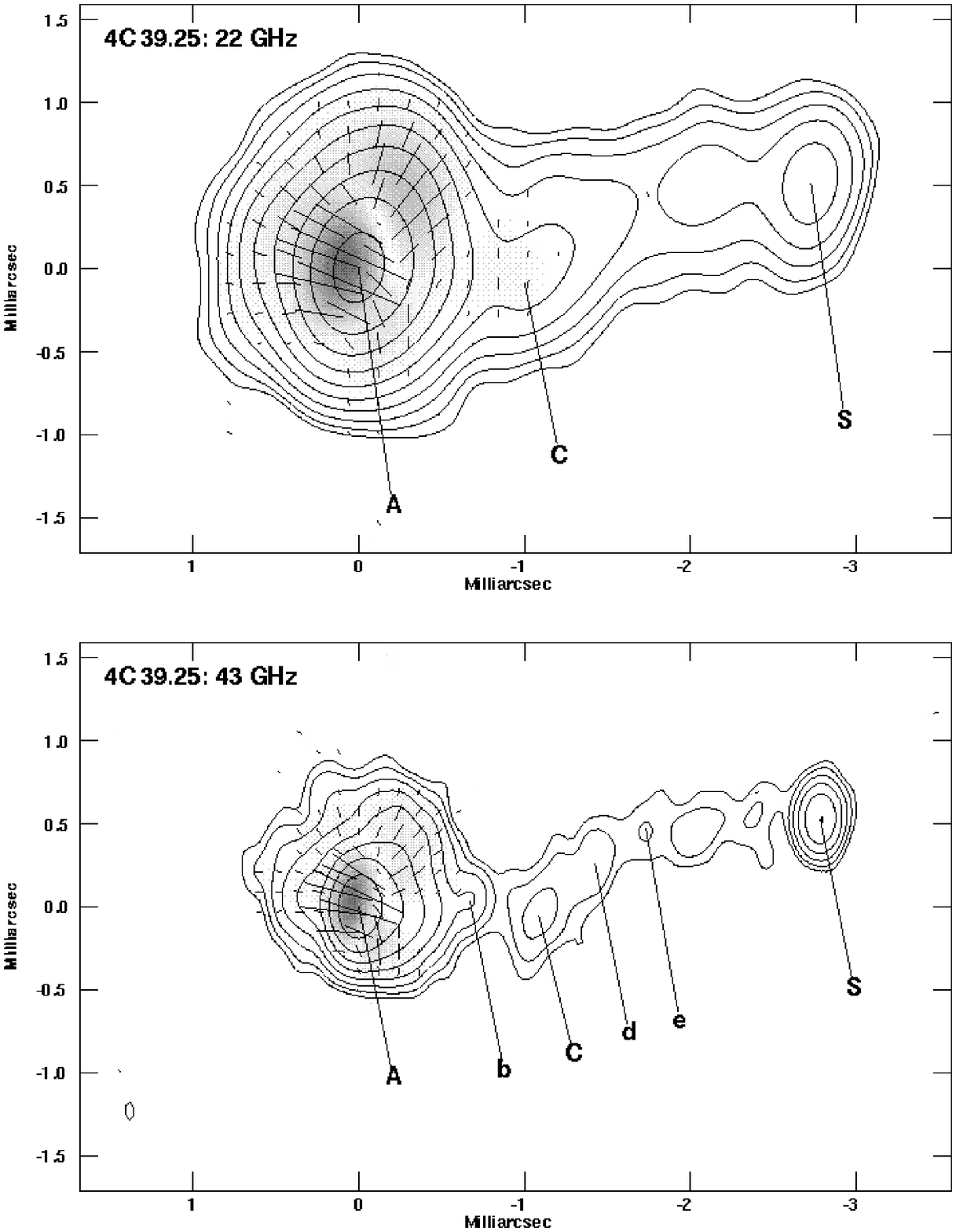}
\caption{VLBA total intensity images of 4C~39.25 at 22
and 43 GHz, epoch 1999.03, with electric polarization vectors superimposed. The
greyscale indicates linearly polarized intensity. }
\end{figure*}

\pagestyle{empty} 
\begin{figure*}
\epsscale{1}
\plotone{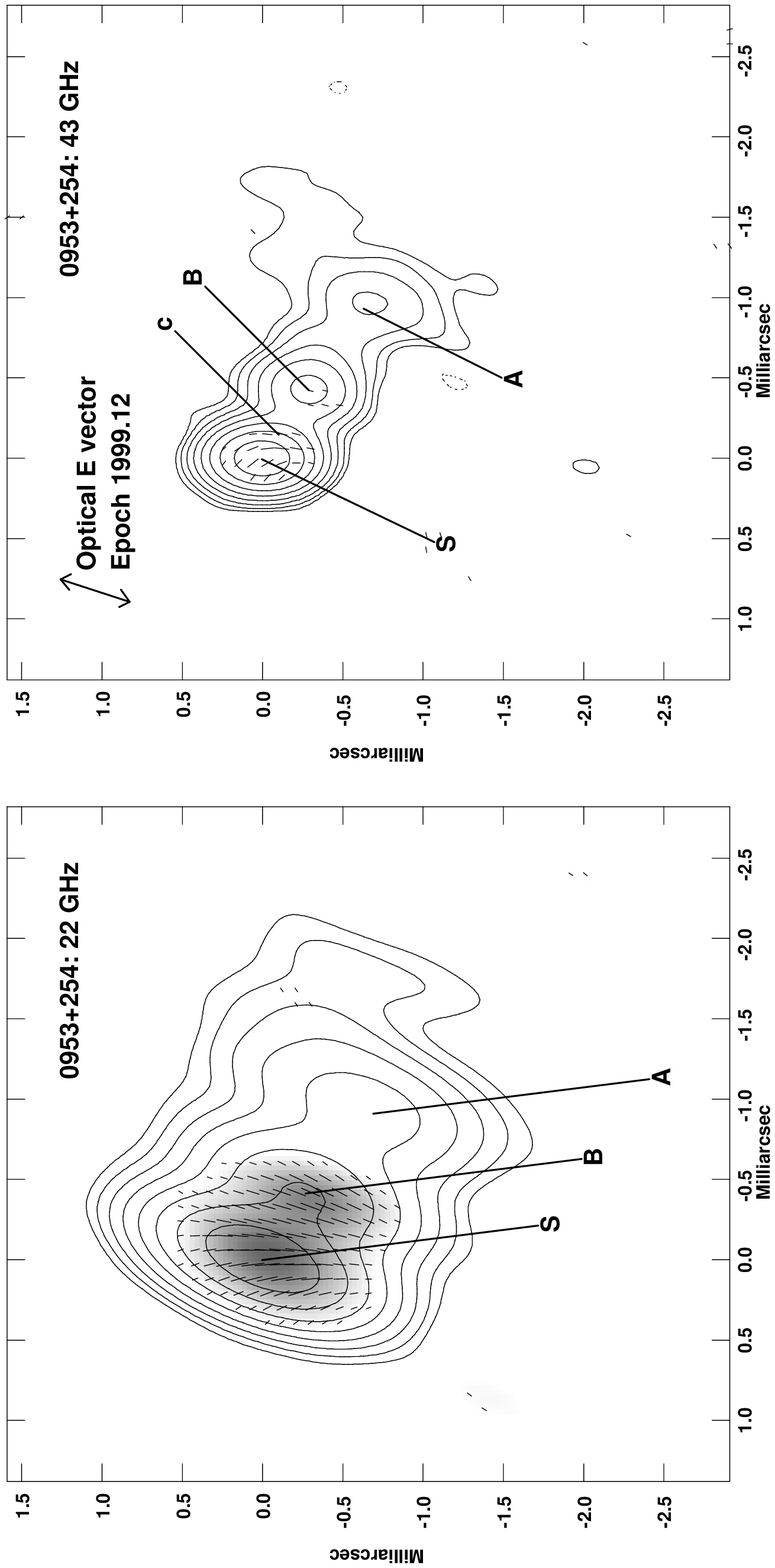}
\caption{VLBA total intensity images of 0953+254 at 22
and 43 GHz, epoch 1999.03, with electric polarization vectors
superimposed. The greyscale in the 22 GHz image indicates linearly
polarized intensity. The position angle of the optical electric vector,
epoch 1999.12, is indicated in the 43 GHz panel. }
\end{figure*}

\begin{figure*}
\epsscale{1}
\plotone{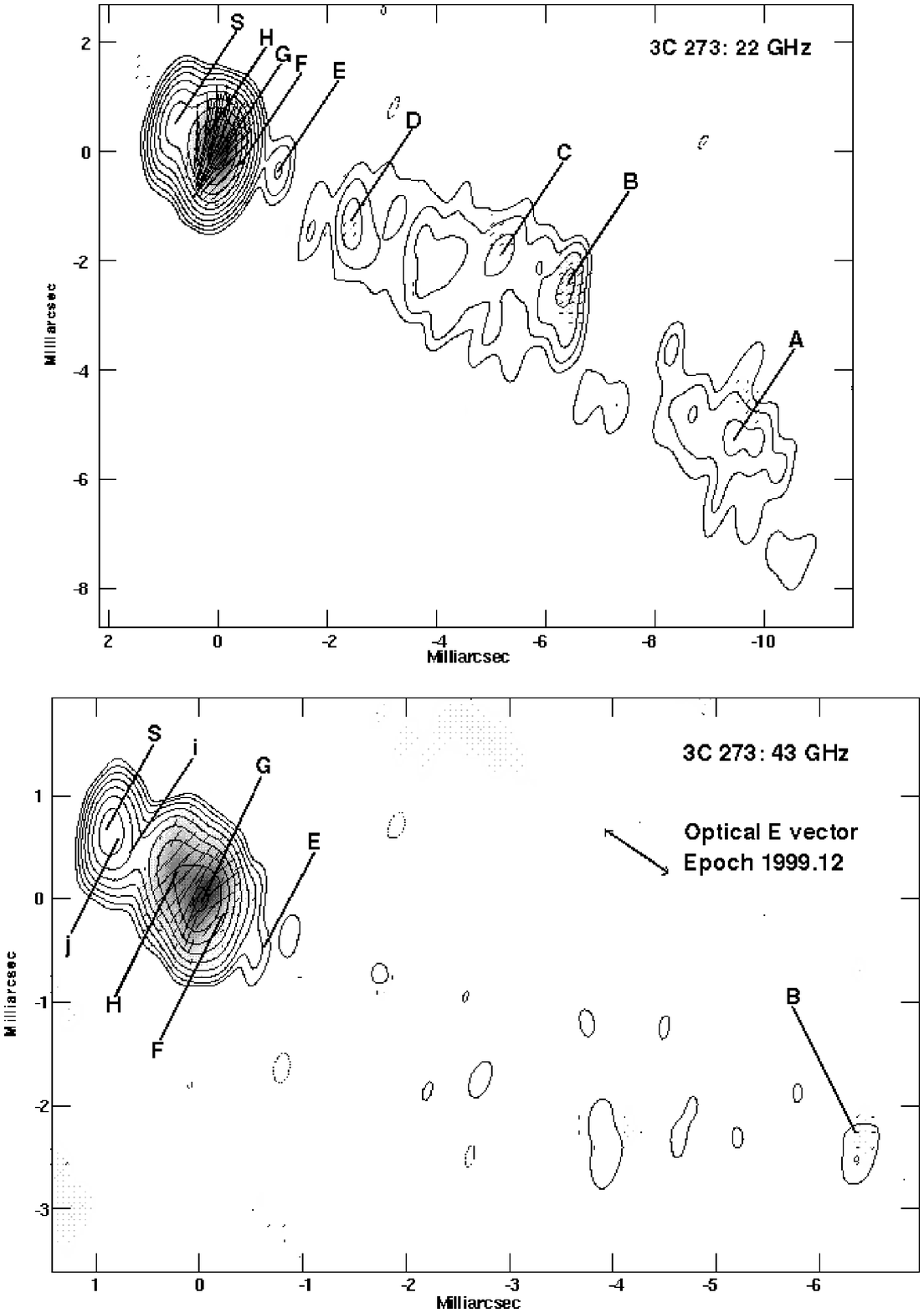}
\caption{VLBA total intensity images of 3C~273 at 22
and 43 GHz, epoch 1999.03, with electric polarization vectors
superimposed. The greyscale indicates linearly polarized intensity
. The position angle of the optical electric vector,
epoch 1999.12, is indicated in the 43 GHz panel. }
\end{figure*}

\begin{figure*}
\epsscale{1}
\plotone{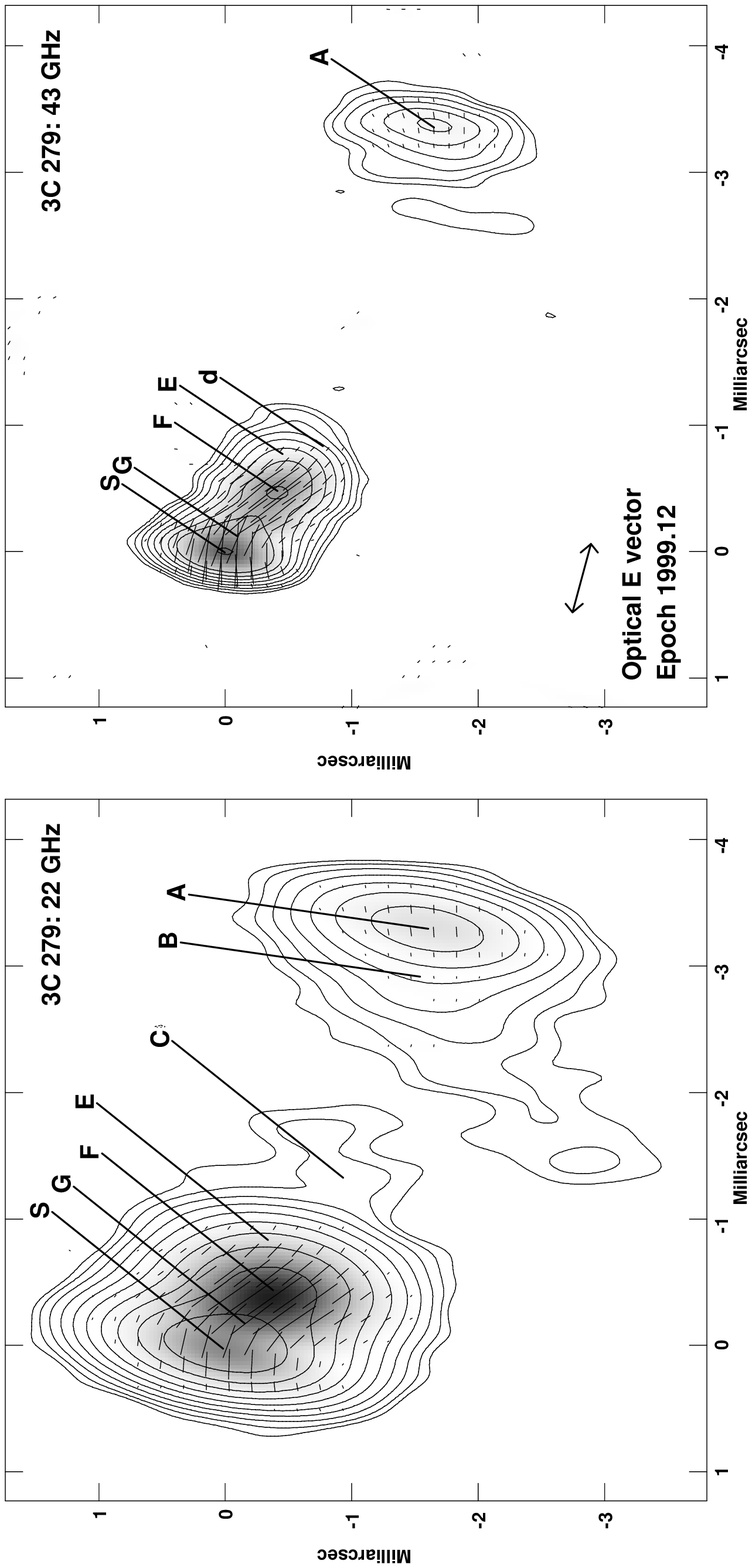}
\caption{VLBA total intensity images of 3C~279 at 22
and 43 GHz, epoch 1999.03, with electric polarization vectors
superimposed. The greyscale indicates linearly polarized intensity. The position angle of the optical electric vector,
epoch 1999.12, is indicated in the 43 GHz panel. }
\end{figure*}

\begin{figure*}
\epsscale{1}
\plotone{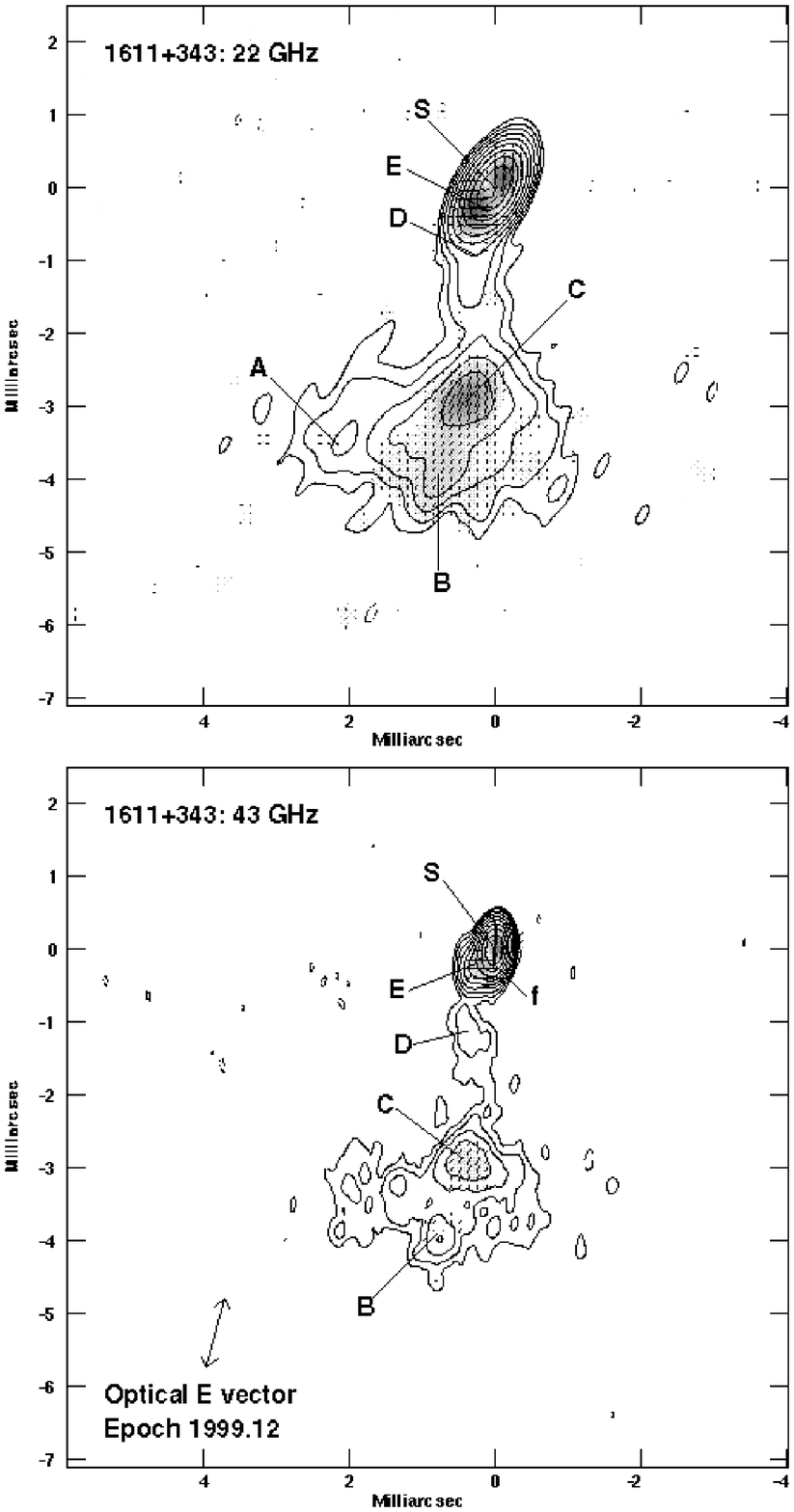}
\caption{VLBA total intensity images of 1611+343 at 22
and 43 GHz, epoch 1999.03, with electric polarization vectors
superimposed. The greyscale indicates linearly polarized intensity
.  The position angle of the optical electric vector,
epoch 1999.12, is indicated in the 43 GHz panel.}
\end{figure*}

\pagestyle{empty} 
\begin{figure*}
\epsscale{1}
\plotone{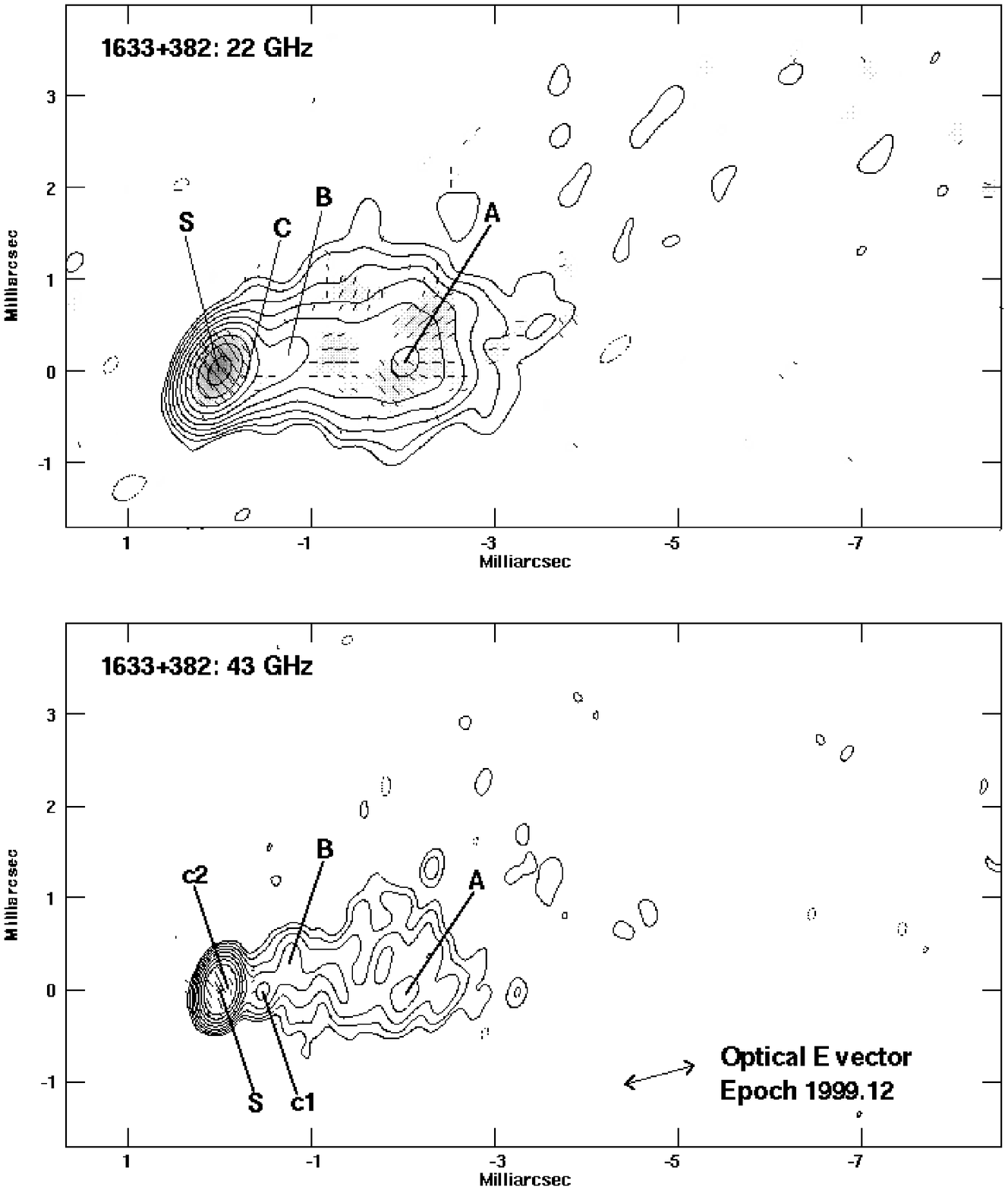}
\caption{VLBA total intensity images of 1633+382 at 22
and 43 GHz, epoch 1999.03, with electric polarization vectors
superimposed. The greyscale in the 22 GHz image indicates linearly
polarized intensity. The position angle of the optical electric vector,
epoch 1999.12, is indicated in the 43 GHz panel. }
\end{figure*}

\pagestyle{empty} 
\begin{figure*}
\epsscale{1}
\plotone{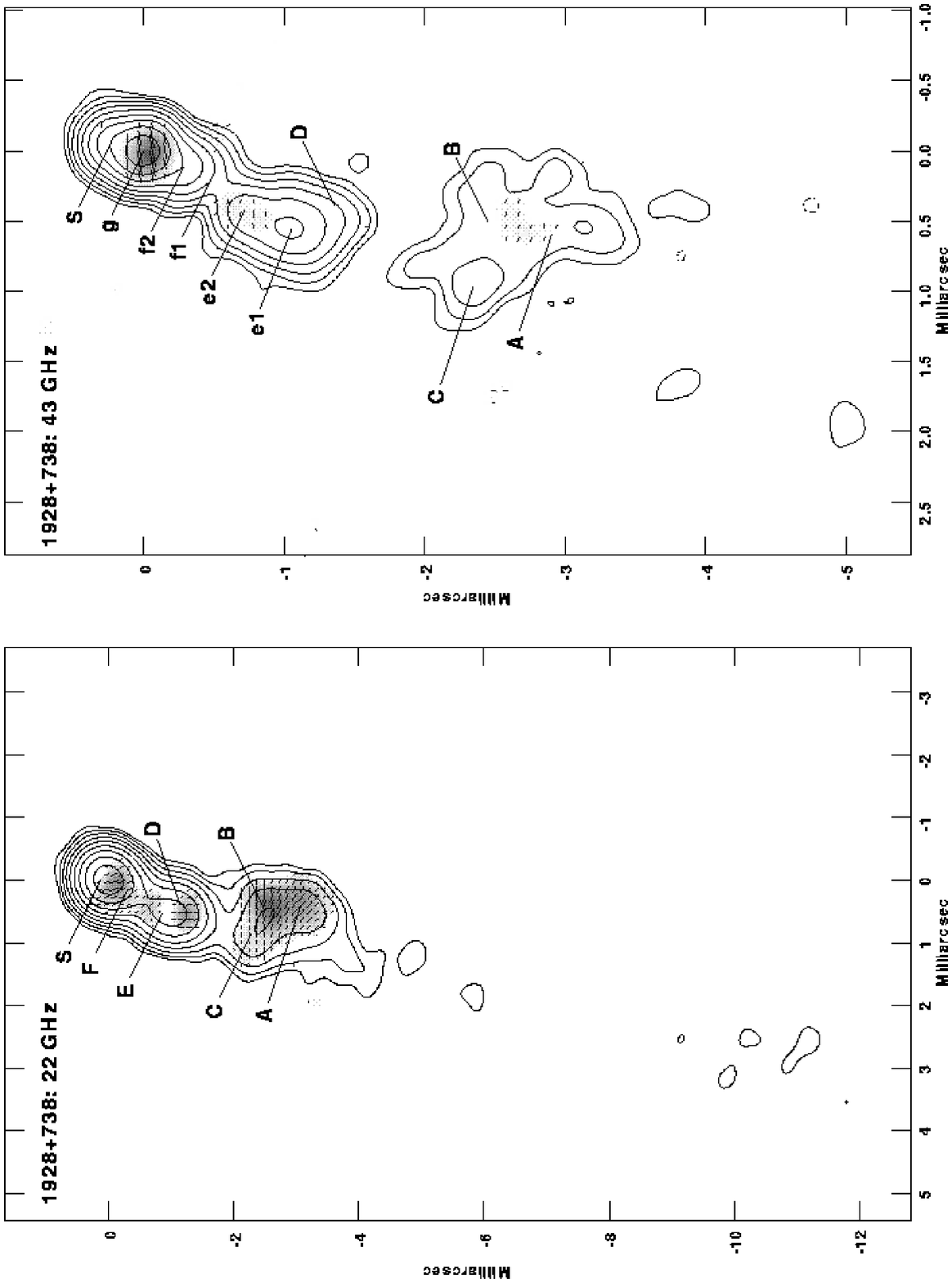}
\caption{VLBA total intensity images of 1928+738 at 22
and 43 GHz, epoch 1999.03, with electric polarization vectors
superimposed. The greyscale indicates linearly polarized intensity.}
\end{figure*}

\begin{figure*}
\epsscale{1}
\plotone{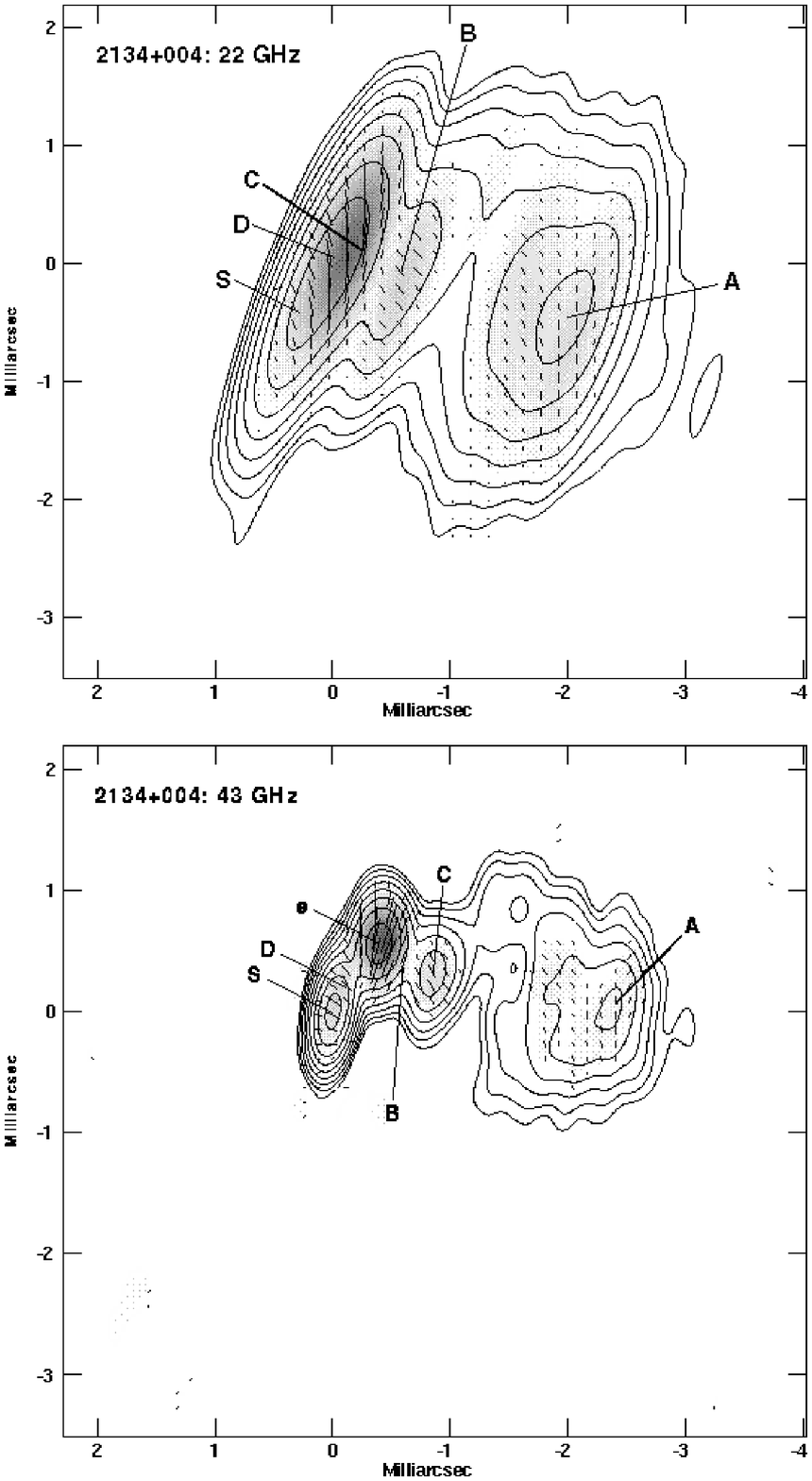}
\caption{VLBA total intensity images of 2134+004 at 22
and 43 GHz, epoch 1999.03, with electric polarization vectors
superimposed. The greyscale indicates linearly polarized intensity.}
\end{figure*}

\pagestyle{empty} 
\begin{figure*}
\epsscale{1}
\plotone{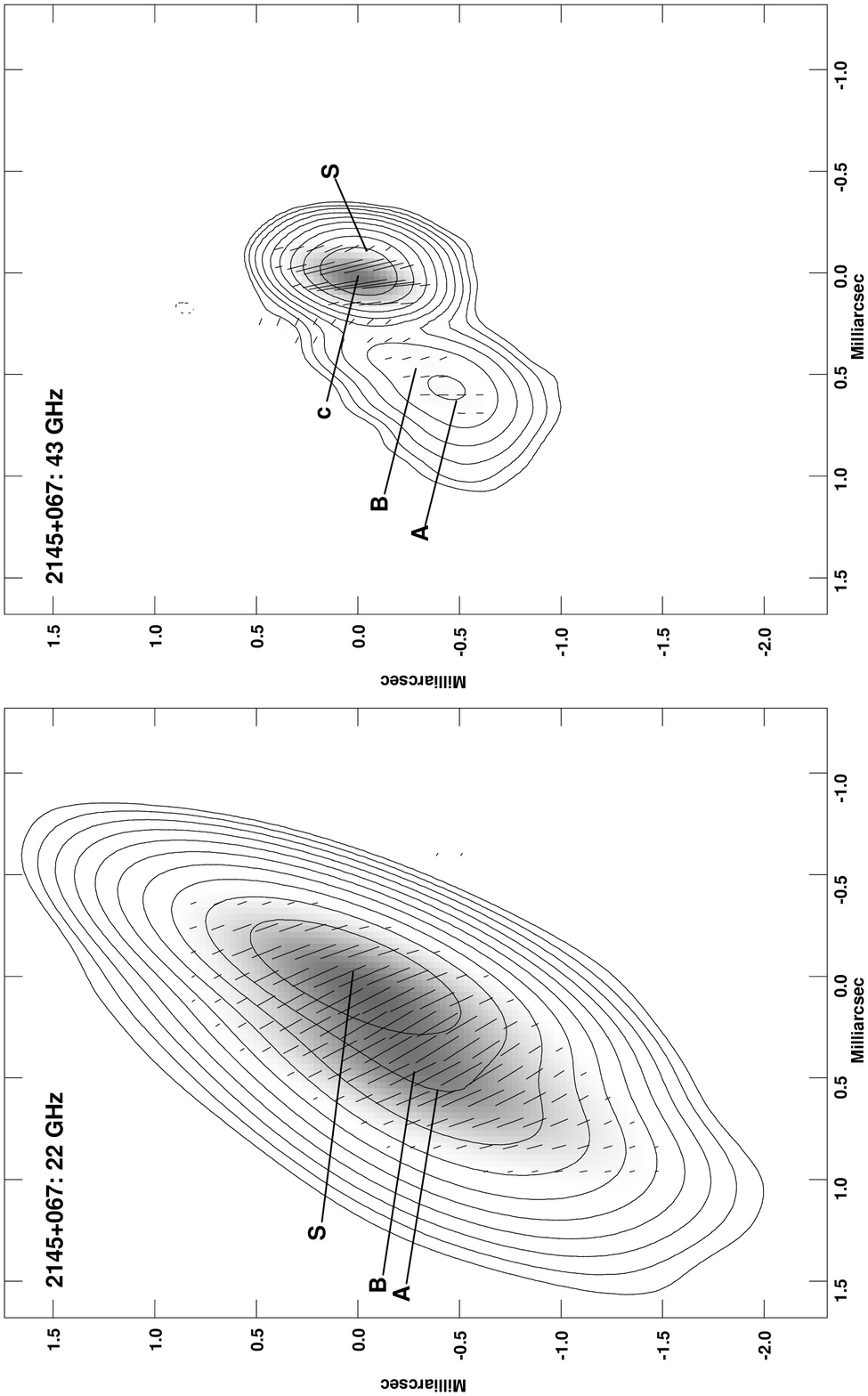}
\caption{VLBA total intensity images of 2145+067 at 22
and 43 GHz, epoch 1999.03, with electric polarization vectors
superimposed. The greyscale indicates linearly polarized intensity.}
\end{figure*}

\begin{figure*}
\epsscale{1}
\plotone{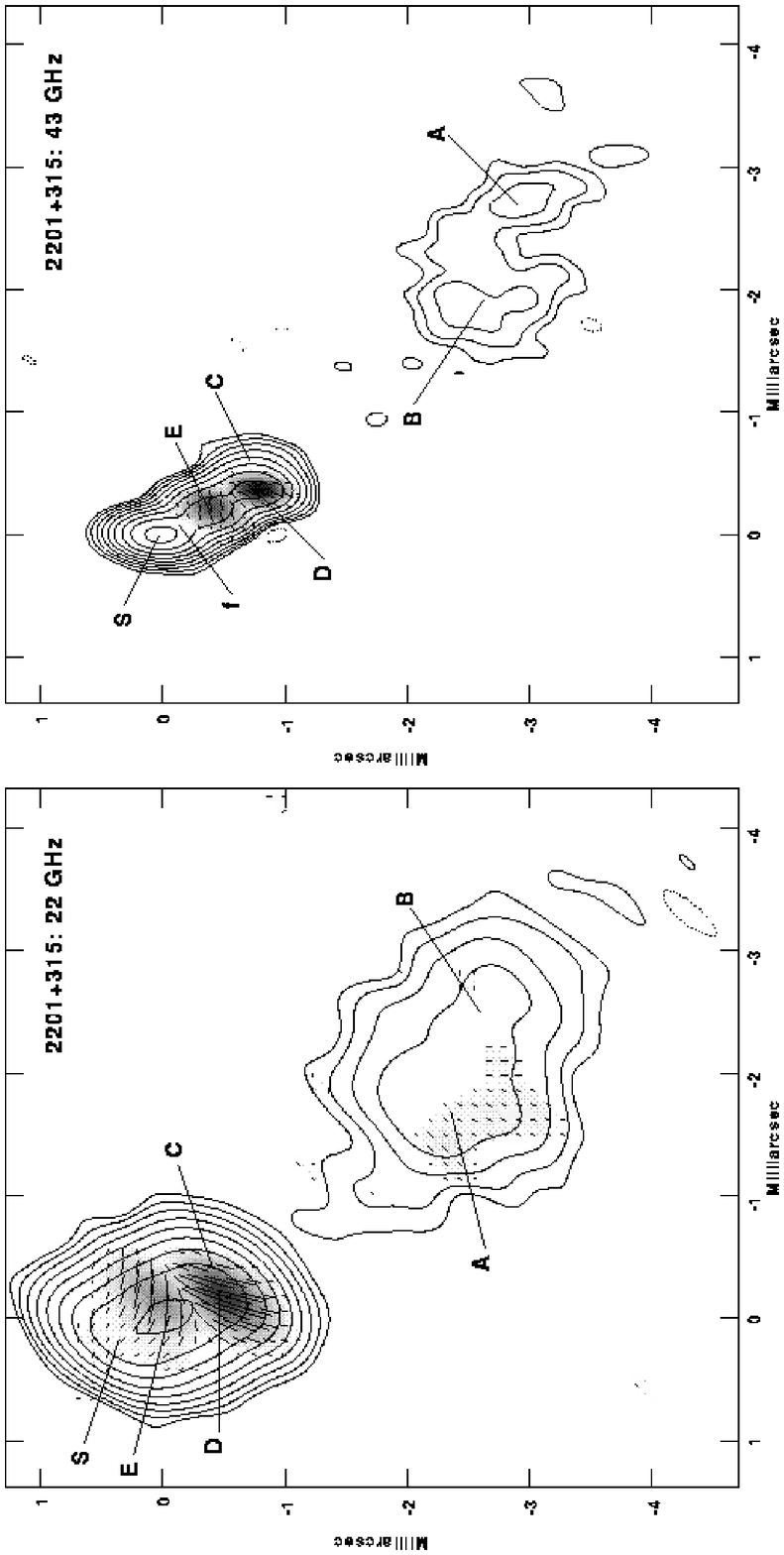}
\caption{VLBA total intensity images of 2201+315 at 22
and 43 GHz, epoch 1999.03, with electric polarization vectors
superimposed. The greyscale indicates linearly polarized intensity.}
\end{figure*}

\begin{figure*}
\epsscale{1}
\plotone{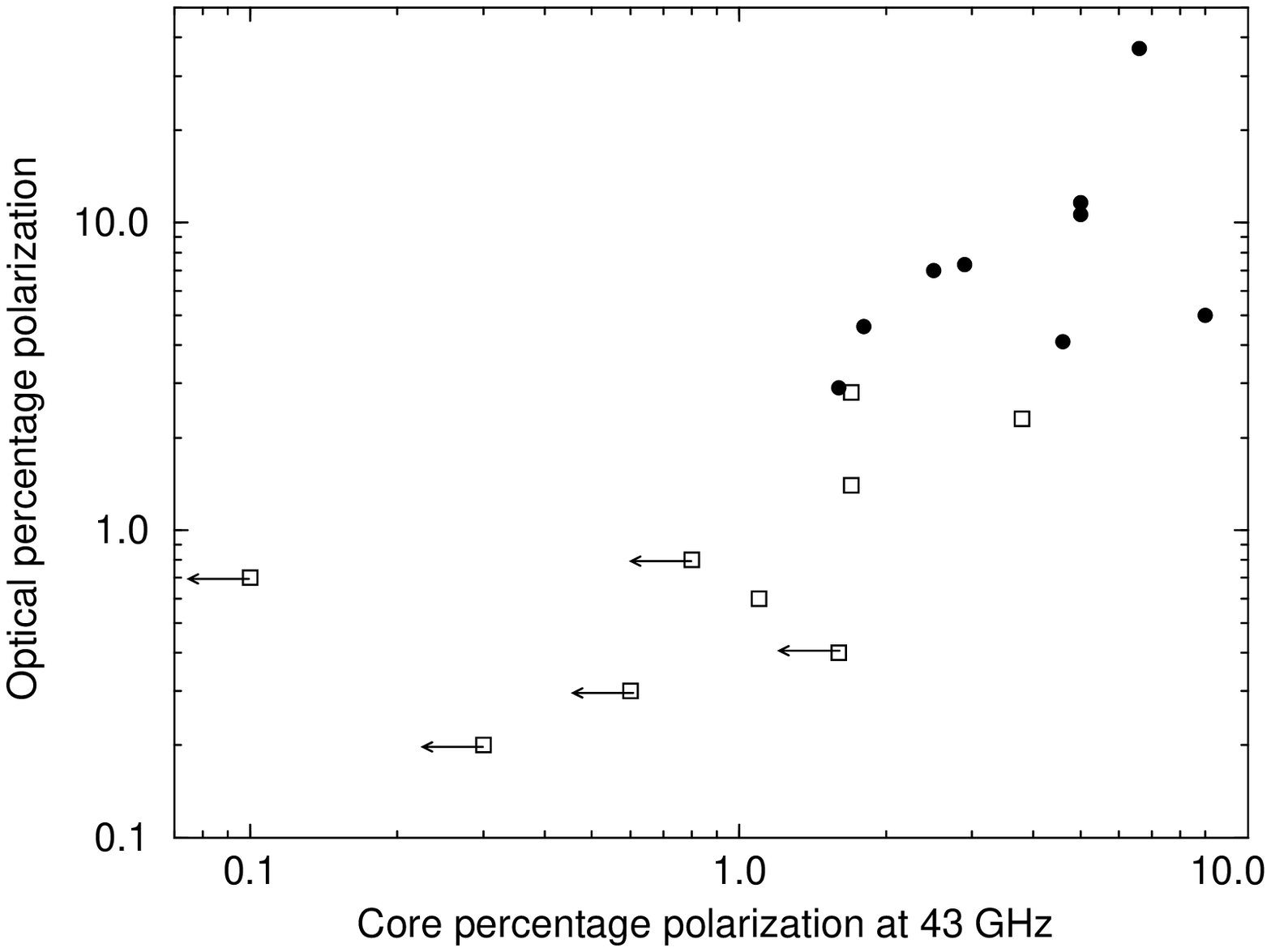}
\caption{\label{core_m_vs_m_opt} Total optical
percentage polarization plotted against that of the unresolved radio
core at 43 GHz. The LPRQs are indicated by the open squares, and the
HPQs by the filled circles. Upper limits are given for sources with no
detected core polarization.}
\end{figure*}

\subsection{Optical observations\label{optical_obs}}
We obtained optical photopolarimetric data on ten of our sample
objects with the Steward Observatory 60-inch telescope at Mt. Lemmon
on UT 1999 February 12-14.  These were taken with the ``Two-Holer''
polarimeter/photometer \citep*{SSS85}, which uses two RCA C31034 GaAs
photomultiplier tubes that are sensitive from 3200--8600 \AA. Our data
acquisition and reduction procedure followed that of
\citet*{SHA92}. All of our optical observations were unfiltered, and
obtained using a $4\arcsec$ circular aperture except for 3C~273, where
an $8\arcsec$ aperture and a Kron-Cousins R filter ($\lambda_c \sim
6400$ \AA) were used. The effective wavelength of the unfiltered
observations is $\sim$6000 \AA, but is dependent on the shape of the
optical spectrum.  The fractional degree of linear polarization
($m_{\rm opt}$) listed in Table 4 has been corrected for Ricean bias
\citep{WK74}.  In the few instances where only an upper limit can be
placed on $m_{\rm opt}$, a 2-sigma limit is listed without any bias
correction.  In these instances, the polarization position angle,
$\chi_{\rm opt}$, is undefined.  No correction to these data was made
for instrumental polarization since Two-Holer yields $< 0.1\%$
polarization for known unpolarized stars.  BD+59$^\circ$389 and
HD155528 were used to calibrate $\chi_{\rm opt}$ \citep{SEL92}.

The observation of 1633+382 on UT 1999 February 14 showed $m_{\rm opt}$
= 7.0$\pm$0.5\%.  By definition, this measurement places the object into the 
HPQ category.  Previous optical polarimetry never showed this object with
$m_{\rm opt} > 3$\% \citep{MS84,IT90,ILT91,WWB92}.

The remainder of our sources could not be observed due to observing
time and sun-angle constraints; for these we list previously published
data from the literature in Table~\ref{coredata}. Wherever possible,
we used values from \citet{IT90}. These authors attempted to limit
observational bias by tabulating only first-epoch polarization
measurements for each source made after 1968. For those sources
without complete polarization data in \citet{IT90}, we used values
from \citet{WWE92} and \citet{ILT91}.

\begin{figure*}
\epsscale{1}
\plotone{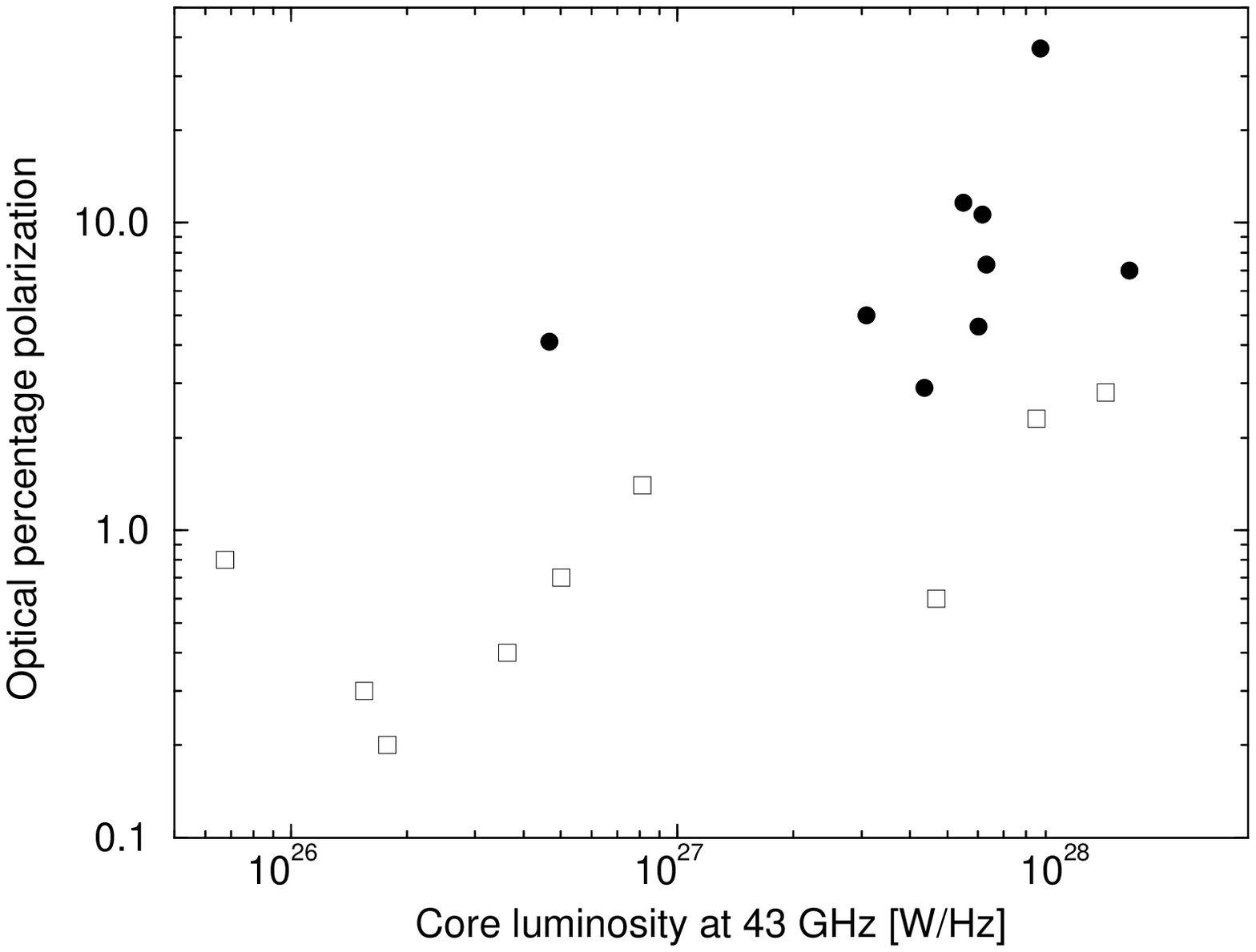}
\caption{\label{m_opt_vs_L_core}Total optical
percentage polarization plotted against luminosity of the unresolved
radio core component at 43 GHz. The LPRQs are indicated by the open squares, and the
HPQs by the filled circles.}
\end{figure*}

\begin{figure*}
\epsscale{1}
\plotone{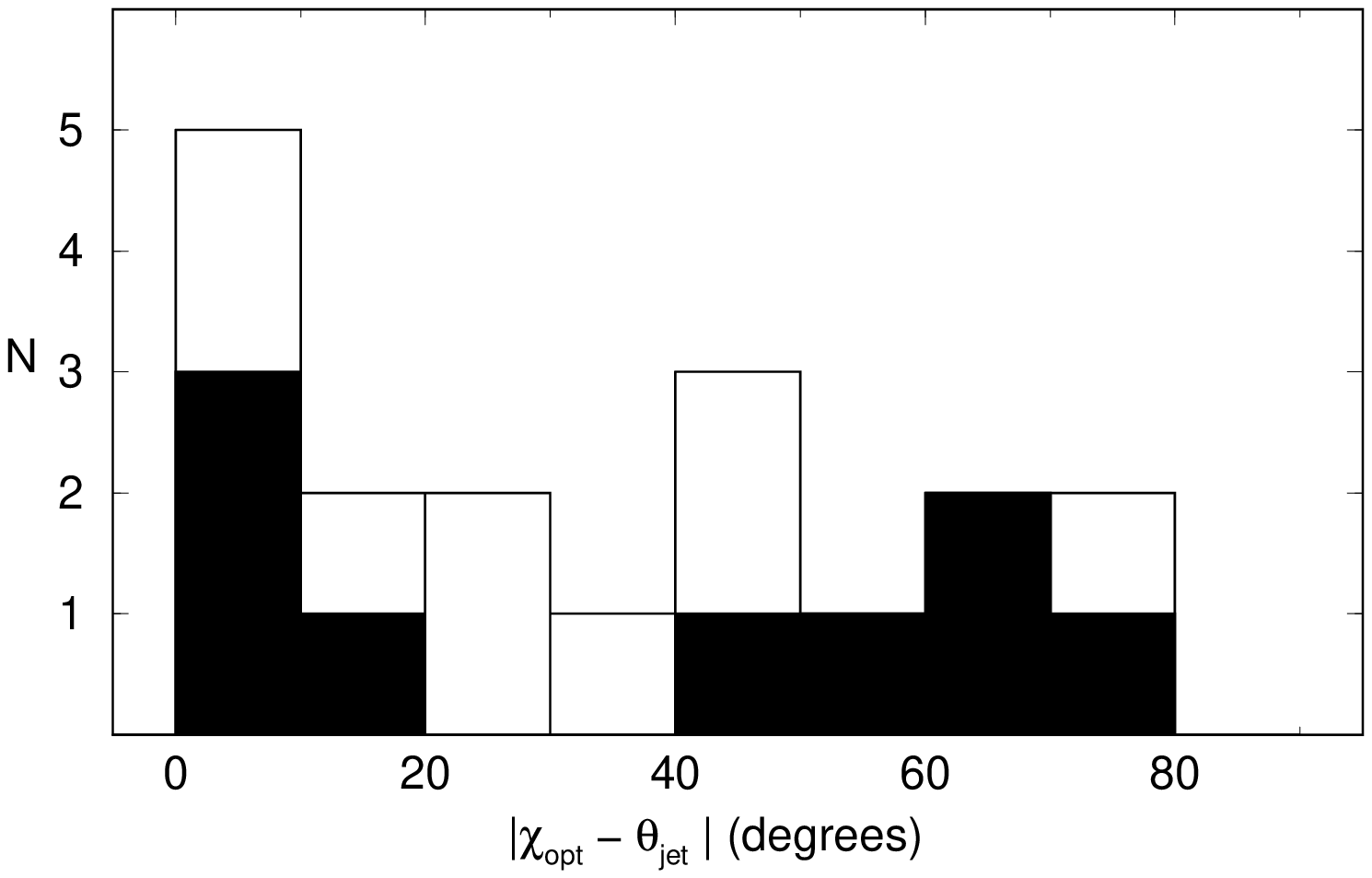}
\caption{\label{dpaopt}Histogram of optical polarization vector
($\chi_{\rm opt}$)
offset with respect to the innermost jet direction, with the LPRQs shaded. Values near
zero indicate a good alignment between the electric polarization
vector and the jet. }
\end{figure*}

\section{Results}

\subsection{Properties of radio core components\label{cores}}
One of the main goals of this study is to characterize the magnetic
field structures of quasars on the smallest scales, and to look for
connections with their optical properties. In this section we discuss
new evidence showing that the optical polarization properties of
quasars are directly related to those of their unresolved radio core
components at 43 GHz.

\begin{figure*}
\epsscale{1}
\plotone{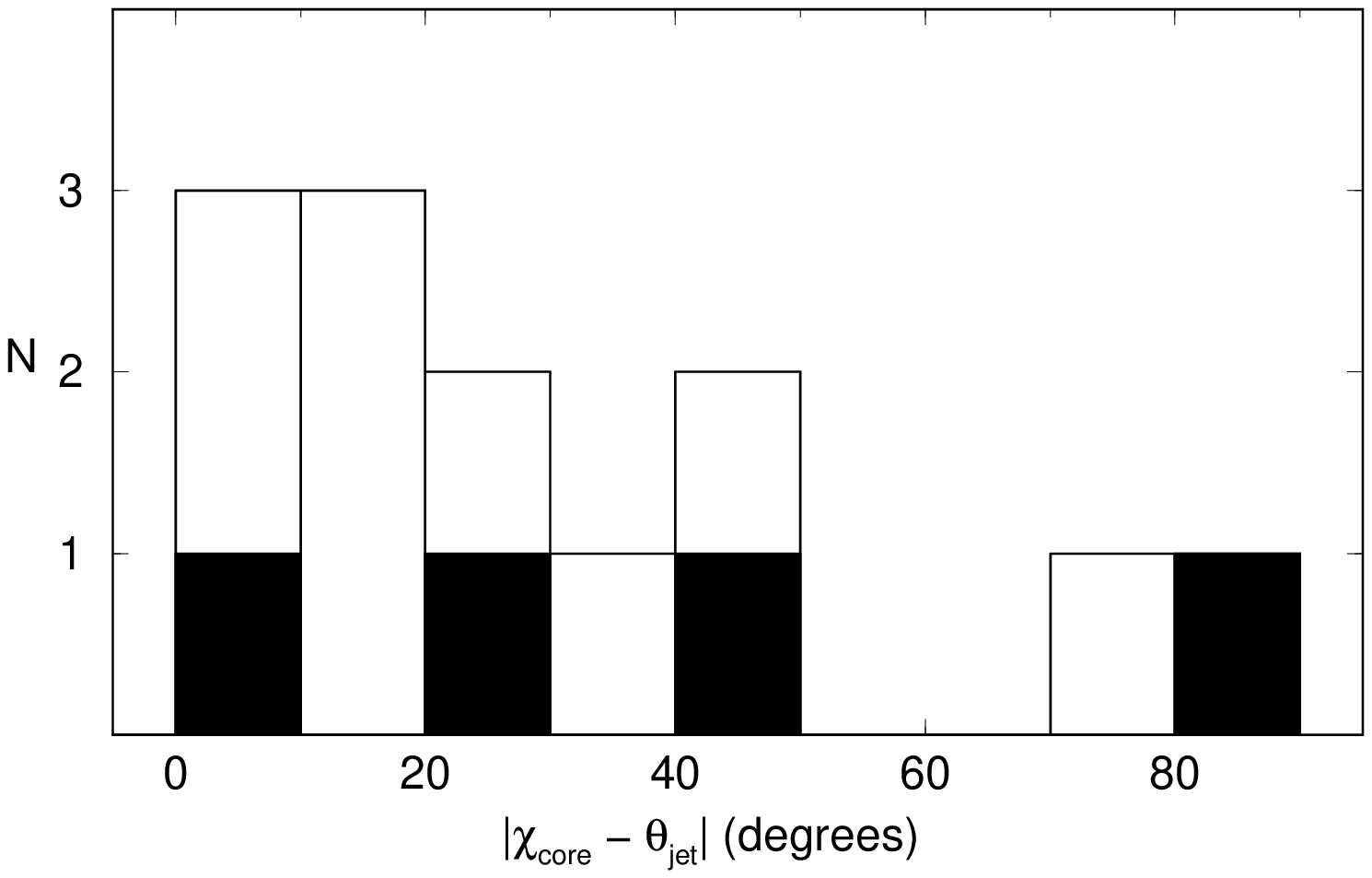}
\caption{\label{dpa_core} Histogram of 43 GHz core electric
polarization vector offsets with respect to the jet direction, with the LPRQs shaded.}
\end{figure*}

\subsubsection{Correlations with optical polarization properties\label{coreopt}}
In Figure~\ref{core_m_vs_m_opt} we show a log-log plot of core
polarization at 43 GHz versus total optical percentage polarization
($m_{\rm opt}$) for our HPQ and LPRQ samples. The relatively small scatter
in this plot suggests that these two quantities are related. To
evaluate the significance of this correlation (and others presented in
this paper), we performed a Kendall's non-parametric rank test, which
uses the relative order of ranks in a dataset to determine the
likelihood of a correlation. A test on these variables indicate that
this correlation is significant at the 99.995\% level.

Previous attempts to find a correlation between radio core and optical
polarization have met with only moderate success, due to a lack of
sufficient sample size and/or spatial resolution. \citet{ILT91} found
a correlation between total (single-dish) polarization at 15 GHz and
optical polarization, which vanished at lower radio frequencies.
Since higher radio frequencies sample emission closer to the base of
the jet, this strongly argued in favor of the optically polarized
emission being emitted close to the core. \citet{GSS96} found a hint of a
possible correlation between 5 GHz radio VLBI core polarization and
$m_{\rm opt}$, but had only seven sources in their sample.
\citet{CWR93} observed a larger sample of 24 AGNs from the
Pearson-Readhead survey at 5 GHz, but detected VLBI core polarization
in only a handful of objects. Recent VLBA observations of these same
objects at 43 GHz by \citet{LPT99} show that some of their
non-detections were likely caused by blending of polarized components
near the core, which tends to lower the overall degree of
polarization. In the case of our sample, 9 of 11 sources had cores at
22 GHz that were in fact a blend of one or more components, as
revealed by our 43 GHz data. Indeed, if we compare optical
polarization with {\it total integrated} polarization at 43 GHz for
our objects, the correlation significance drops to 83\%.

We find the level of optical polarization to be positively correlated
with several other source properties at the 95\% level or
higher. These are 43 GHz core luminosity, total spectral index between
43 and 22 GHz, and core dominance ($R$). The latter quantity is
defined as the ratio of core to extended (parsec-scale) flux at 43
GHz. \citet{IT90} found a similar correlation between $m_{\rm opt}$
and $R$ for the K\"uhr 2 Jy sample of radio sources but found no
correlation between $R$ and total integrated polarization at 5 GHz. We
obtain a similar null result for $R$ and {\it total} polarization
level at 43 GHz, but find that $R$ is positively correlated with the
{\it core} polarization level at the 99.6\% confidence level.

These correlations all indicate that the more optically polarized
quasars have brighter radio cores at 43 GHz, as a bright core tends to
flatten the spectral index and increase $R$.  We plot optical
polarization versus 43 GHz core luminosity for our samples in
Figure~\ref{m_opt_vs_L_core}.

\subsubsection{Electric vector position angles\label{coreEVPAs}}
Despite the strong correlation between optical and 43 GHz core
polarization, we find no correlation between the electric vector
position angles ($\chi$) at these two wavelengths (observed
approximately a month apart) for the entire dataset. A similar null
result was found at lower radio frequencies by
\citet{RS85} in their analysis of a larger sample of radio-loud
quasars. It is unlikely that the electric vectors are significantly
affected by Faraday rotation at 43 GHz, given the typical nuclear
rotation measures of quasars seen in this sample and others ($\lesssim
3000\; \rm \ rad \ m^{-2}$; see \S5).  A more likely explanation is
that the polarization vectors have rotated in the time interval
between the optical and radio measurements. This would imply that the
$\chi$'s of blazars are significantly more variable than their
fractional polarizations. Indeed, rapid swings in $\chi$ are not
uncommon in blazars \citep[e.g.,][]{AAH99}, with changes sometimes
occurring on timescales $\lesssim 2$ weeks \citep{GKK98}.  Another
possibility is that the optically polarized emission may be
originating in a strongly shocked component in the jet, instead of the
core. \citet{GSS96} found evidence for this in a small sample of BL
Lacertae objects observed at 5 GHz.

If we restrict our analysis to those sources in our sample with
near-simultaneous optical and radio observations, we find similar
alignments of $\chi_{\rm opt}$ with those of jet components in four of
five sources (3C 279, 3C273, 1611+343, and 0953+254) at 43 GHz. The
only source with similar 43 GHz {\it core} component $\chi$ and $\chi_{\rm
opt}$ is 1611+343. Due to the large range of $\chi$ seen in the jets
of some sources and the small number of sources, it is possible that
these alignments are merely due to chance. A proper study of
radio/optical $\chi$ alignments requires truly simultaneous optical
and radio observations over a sufficient timescale to determine
whether changes in the optical $\chi$'s follow those seen in the radio
jets, or the unresolved core components.

Many previous VLBI polarization studies have compared the optical and
radio $\chi$'s to the jet direction near the core
($\theta_{jet}$), as the simplest form of the shock-in-jet model
predicts that a strong transverse shock will preferentially strengthen
the perpendicular component of an initially tangled magnetic field. In
the absence of any Faraday rotation, the electric vectors of the
shocked region will be rendered parallel to the jet. In
Figure~\ref{dpaopt} we plot the distribution of $|\chi_{\rm opt} -
\theta_{jet}|$ for our samples.  This quantity represents the
difference between the optical $\chi$ and the direction of the jet
closest to the core. We define $\theta_{jet}$ as the position angle of
the innermost jet component with respect to the core. Due to the
$180\arcdeg$ ambiguity in the measured polarization position angles,
$|\chi_{\rm opt} - \theta_{jet}|$ can never exceed $90\arcdeg$. The
overall distribution is peaked at zero, where the optical $\chi$'s are
parallel to the jet.  This is in agreement with the results of
previous studies \citep{I87,R90,ILT91}.  The distribution for the
LPRQs (shaded) is possibly bi-modal, with peaks at zero and roughly
60\arcdeg, which was also seen by \citet{ILT91} in a much larger
sample.

We plot an analogous distribution for the radio core $\chi$'s at 43 GHz in
Figure~\ref{dpa_core}. Only four LPRQ cores are shown in this plot,
since the remainder did not have any detectable polarized flux at 43
GHz. Again the distribution is peaked near zero, and only two sources
have core magnetic fields that are roughly parallel to the jet
($|\chi_{\rm opt} - \theta_{jet}| \gtrsim 70\arcdeg$). We find that all of
the highly polarized cores with $m_{43} > 3\%$ have core $\chi$'s aligned
to within $30\arcdeg$ of the local jet direction. These findings support
the scenario described in \S\ref{coreopt}, in which the core
polarization in these sources originates in a shock near the base of
the jet.

\begin{figure*}
\epsscale{1.0}
\plotone{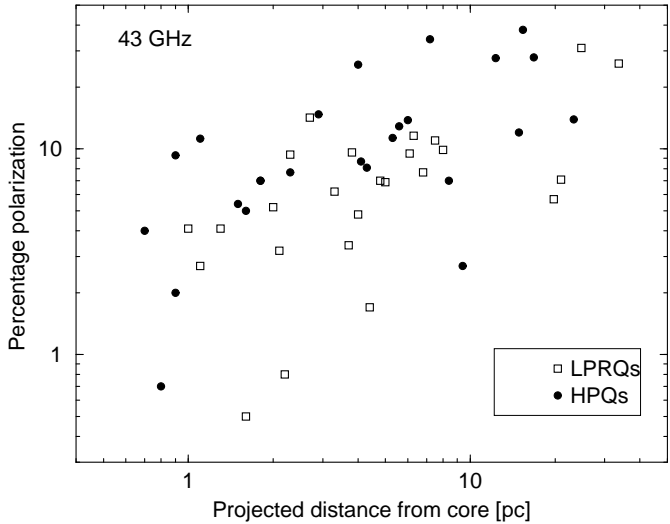}
\caption{\label{m43_vs_r43}Percentage polarization
of jet components at 43 GHz plotted against projected distance from
the core in parsecs.}
\end{figure*}

\subsection{Jet properties}

The steep radio spectral indices of AGN jets ($\alpha \sim -0.8$)
generally cause them to drop below the sensitivity level of current
high-frequency VLBI images at distances of more than a few
milliarcseconds from the core. Nevertheless, the large improvement in
angular resolution and reduced source opacity at high frequencies
allow for a more thorough investigation of the nature of AGN jets
close to the core. In this section we discuss the overall jet
polarization properties of our combined HPQ and LPRQ samples.

The strongest trend we find in the inner jets of our sample is an
steady increase in fractional polarization as you move downstream from
the core. There is no significant decrease in total intensity with
core distance, however, as was seen in the sample of \citet{CWR93} at
5 GHz, which covered regions farther down the jet ($\sim 10-90$ pc,
projected).  Figure~\ref{m43_vs_r43} shows an increase in $m_{43}$
with distance down the jet, for both LPRQs and HPQs. A similar trend
exists in the 22 GHz data. A Kendall's tau test shows these variables
are correlated at a significance level of $\sim 99.99\%$. This
increase in polarization with distance was also seen in quasars (but
not in BL Lac objects) at 5 GHz by \citet{CWR93}, who attributed it to
interaction of the jet boundary with the external medium. According to
these authors, the jet material experiences shear as it moves down the
jet, which organizes its (initially random) magnetic field, and
increases the degree of polarization. We note that the limited dynamic
range of our images does not allow us to measure the polarization of
the underlying jet in most sources, however, but rather that of
polarized knots associated with shocks.

\begin{figure*}
\epsscale{1.0}
\plotone{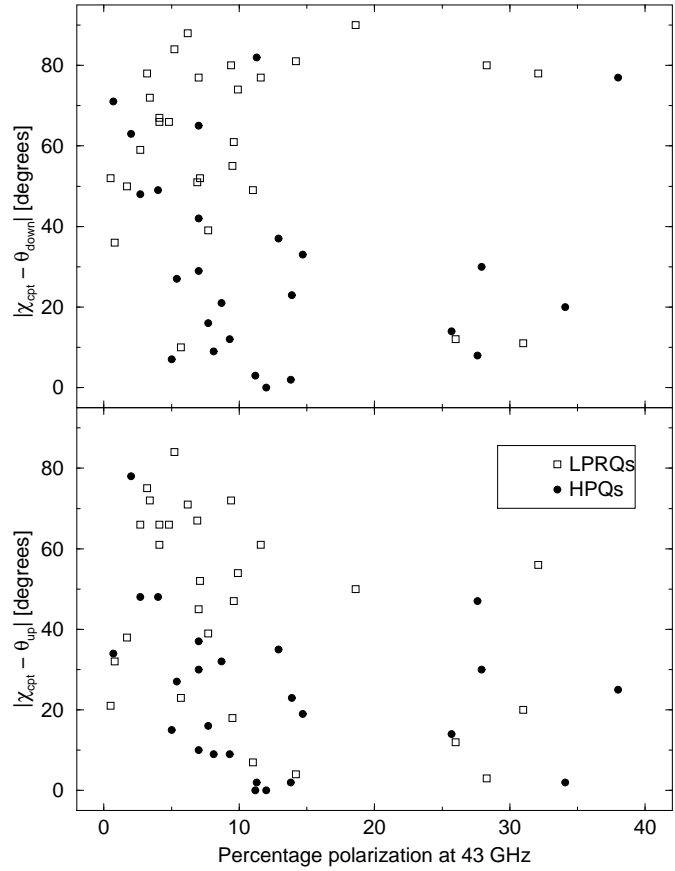}
\caption{\label{m43_vs_DPA43} Top panel: electric polarization vector offset with
respect to the local downstream jet direction plotted against fractional
polarization, for polarized jet components. Bottom panel: same plot
for upstream jet directions. }
\end{figure*}

\begin{figure*}
\epsscale{1.0}
\plotone{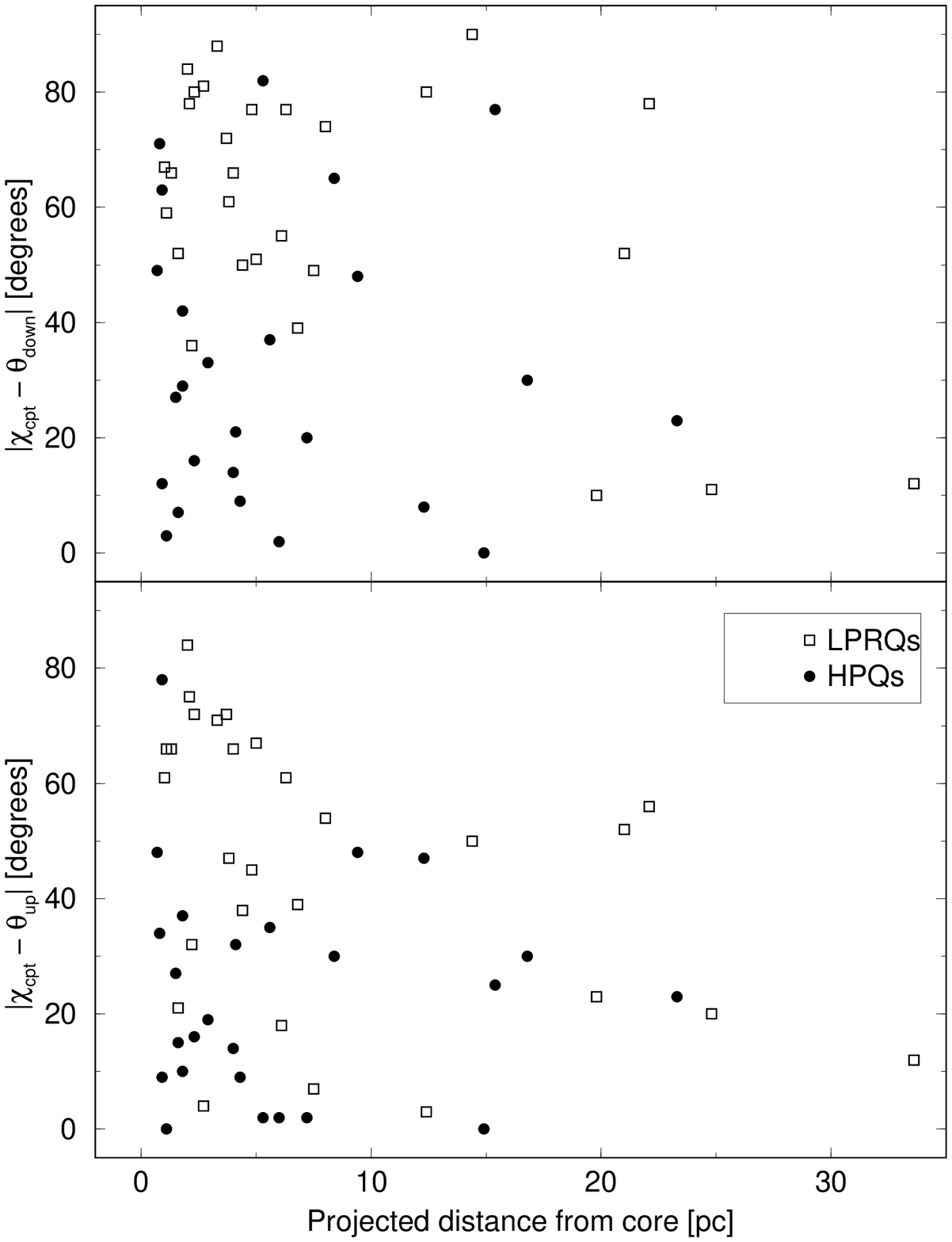}
\caption{\label{r43_vs_DPA43}  Top panel: electric polarization vector offset with
respect to the local downtream jet direction plotted against projected
distance from the core in parsecs. Bottom panel: same plot
for upstream jet directions. }
\end{figure*}
We performed a similar analysis to \S\ref{coreEVPAs} by
comparing the $\chi$ of each polarized component to the local jet
direction. Because many of the jets in our sample undergo significant
bends, the latter quantity is ambiguous in some cases. We therefore
list two local jet directions for each component in
Table~\ref{components}: $\theta_{\rm up}$ represents the direction of the
jet ridgeline upstream of the component, and $\theta_{\rm down}$
represents the jet direction downstream. In cases where nearby jet
emission was weak or absent, we used the position of the nearest
upstream and downstream components to establish the local jet direction. 

We find an anticorrelation (Kendall's tau = $-0.315$; 99.9\%
significance) between fractional polarization and $|\chi_{\rm
cpt}-\theta_{\rm up}|$ (lower panel of Fig.~\ref{m43_vs_DPA43}), where
$\chi_{\rm cpt}$ is the electric vector position angle of the
component at 43 GHz. There is a loosely defined upper envelope to this
distribution, such that there are no highly polarized components with
magnetic fields parallel to the jet. We find no significant
correlation (Kendall's tau = $-0.117$; 78.3\% significance) using the
downstream position angles, plotted in the top panel of
Fig.~\ref{m43_vs_DPA43}.

We also find an upper envelope in the plot of
$|\chi_{\rm cpt}-\theta_{\rm up}|$ versus projected core distance (lower panel
of Fig.~\ref{r43_vs_DPA43}).  All of the components with magnetic
fields nearly parallel to the jet are located very close to the
core. Furthermore, these components are all associated with LPRQs
(with the exception of component c2 of 1633+382, a quasar that would
have been classified as an LPRQ prior to our optical observation on UT
1999 February 14). 

The increasing longitudinal field model of \citet{CWR93} has
difficulty accounting for the trends in Figs.~\ref{m43_vs_DPA43} and
\ref{r43_vs_DPA43}. If we assume that the viewing angle and shock
strengths are constant along the jet, the $\chi$'s should become more
perpendicular to the jet with increasing distance from the core. This
is the opposite of what is seen. Also, this model predicts that the
fractional polarization of components with $\chi$'s parallel to the
jet should not be appreciably lower than those with perpendicular
orientations, which is not the case. For these reasons, we do not
believe the trend of increasing fractional polarization with distance
is caused by an underlying longitudinal magnetic field of increasing
strength. More likely, this trend is due to either an increase in
shock strength with distance, or jet curvature. Although shock
strengths are generally difficult to estimate without good
determinations of shock speed and jet orientation, a test of the
curvature scenario may be relatively straightforward.  

\begin{figure*}
\epsscale{1.0}
\plotone{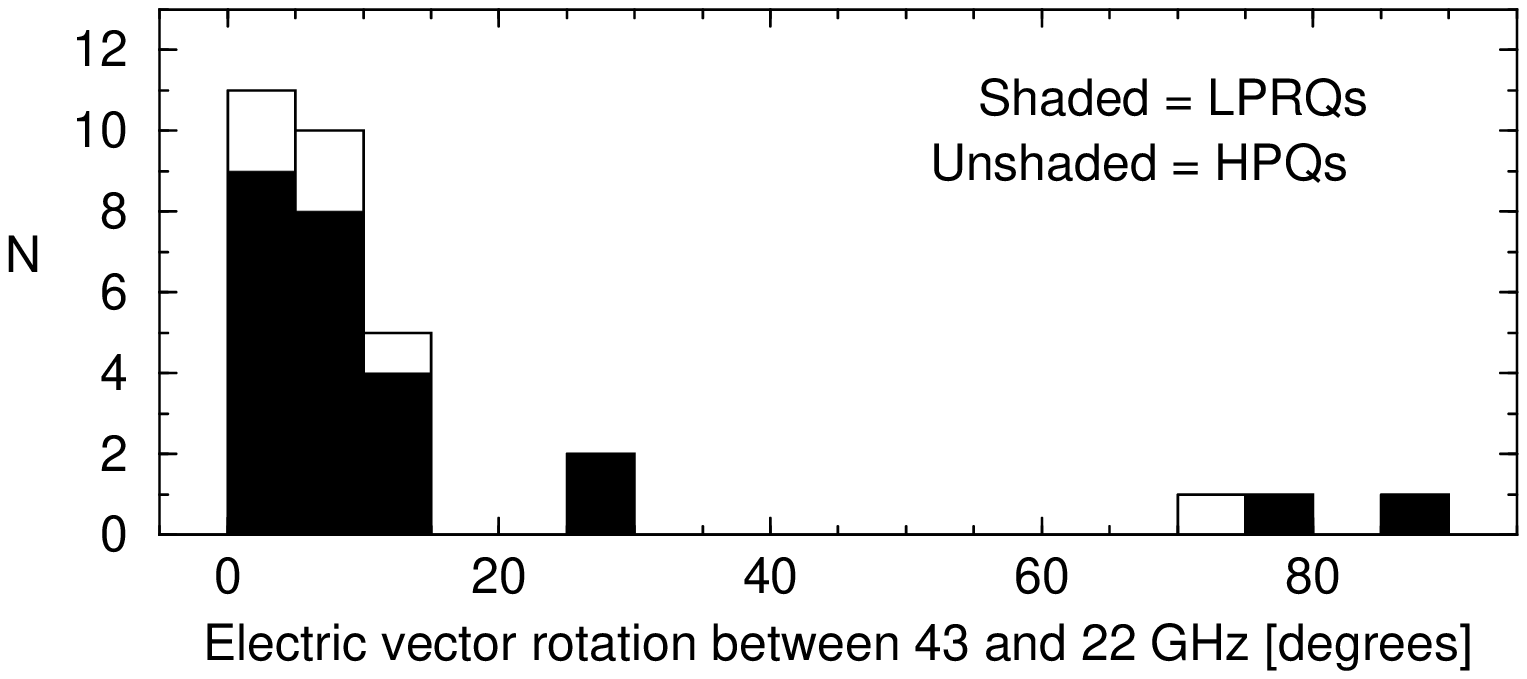}
\caption{\label{EVPArot} Distribution of polarization vector rotation
between 22 and 43 GHz for polarized jet components.}
\end{figure*}

Given the high luminosities of the core components of our sample, it
is likely that they are associated with the most highly aligned
portions of the jet, and are seen at very small viewing angles. If the
jets have simple monotonic bends, they will then be curving away from
the line of sight. This appears to be the case for the gamma-ray
blazars detected by EGRET \citep{MMM00}. According to the simple
transverse shock model \citep{HAA85}, the polarization of a bent jet
should increase along the jet as the viewing angle increases from
originally small values, due to aberration. Even if the bends are not
monotonic, the projected core distance will not increase very rapidly
in those portions of the jet which are bending toward us, thereby
preserving the trend seen in Fig.~\ref{m43_vs_r43}. This bending model
could be tested, for example, by simultaneously measuring the
inter-knot polarization levels and superluminal speeds along the jet,
as these are both affected by orientation in a predictable manner. We
will return to the role of orientation and shock strength in our
discussion of HPQ and LPRQ properties in \S6.

\begin{figure*}
\epsscale{1.0}
\plotone{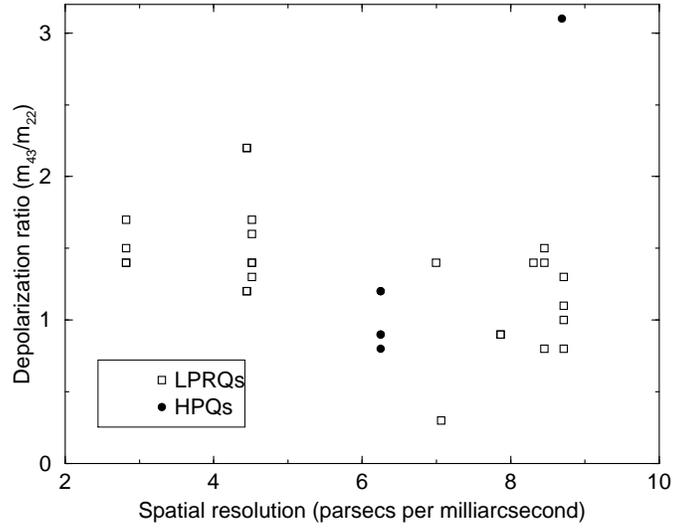}
\caption{\label{depol_vs_angscale}Ratio of
percentage polarization at 43 and 22 GHz for jet components plotted against the spatial
resolution of the restoring beam.}
\end{figure*}

\begin{figure*}
\epsscale{1.0}
\plotone{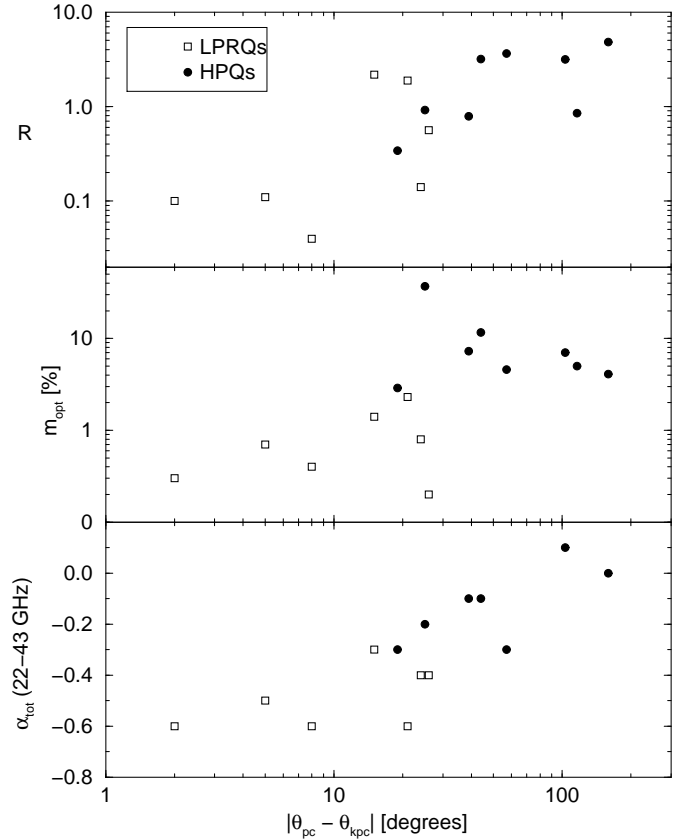}
\caption{\label{bend_corr}Plots of various
source quantities against jet misalignment between
parsec and kiloparsec scales. In top panel is the ratio of core
to extended flux at 43 GHz, the middle panel shows optical percentage
polarization, and the lower panel shows integrated spectral index
between 22 and 43 GHz. }
\end{figure*}

\section{Comparison of magnetic field properties at 22 and 43 GHz\label{Frot}}

Based on the typical integrated rotation measures of AGNs at
cm-wavelengths ($\lesssim 200$ \rotm: \citealt{W93,O89}), we
would expect to see very few differences in the polarization properties of
our sample at 22 and 43 GHz, with the exception of some spatial
resolution effects.  Recent VLBI polarization studies
\citep[e.g.][]{T98,UTP97} have shown, however, that on parsec-scales
the measured rotation measures are generally much higher ($> 1000 $
\rotm).  The most likely origin of these high RMs is external Faraday
rotation by gas in the narrow-line region \citep{T98}. In this section
we discuss the evidence for strong depolarization and Faraday rotation
effects in our LPRQ sample.

\subsection{Core components}
In columns (13) and (14) of Table~\ref{coredata} we list the rotation
in $\chi$ and ratio of percentage polarization at 43 GHz to that at 22
GHz for the core components in our sample. Since several of the cores
had no detectable polarization at 43 GHz, we tabulate data for six
cores only. The depolarization ratios lie between 0.5 and 2.3, while
the $\chi$'s are rotated between 2 and 19 degrees. If we assume that
these $\chi$'s at 22 and 43 GHz follow a standard $\lambda^2$ law, we
can convert the latter into rest-frame rotation measures using
$RM_{rest} = 130.19\; \Delta\chi (1+z)^2 \ \rm rad \ m^{-2}$, where
$\Delta\chi$ is the $\chi$ shift in degrees. Given the uncertainty in
our absolute $\chi$ correction ($\lesssim 5^o$), this yields rotation
measures between 0 and $\sim 6000 \ \rm rad \ m^{-2} $, which are
consistent with values found by \citet{T98} for core-dominated
quasars.

We find a tendency for sources with flatter {\it total} (i.e., single
dish) spectral indices between 1.4 and 5 GHz to have smaller
$m_{43/22}$ ratios (99.5\% confidence according to Kendall's
tau). Since the total spectral index tends to be steeper for sources
with weaker cores, this may indicate the presence of both Faraday and
opacity effects in the core region at longer wavelengths. This opacity
makes it difficult to interpret trends involving the $m_{43/22}$
ratio, as we may simply be sampling spatially distinct regions of
different intrinsic polarization at these two
wavelengths. Observations at an intermediate frequency are required to
investigate this in more detail.

\subsection{Jet components}
In Figure~\ref{EVPArot} we show the distribution of $\Delta\chi =
|\chi_{43}-\chi_{22}|$ for the polarized jet components in our AGN
sample. Although the majority have $\Delta\chi < 15^o$, there are
three with substantial $\chi$ rotations. These components are all very
weakly polarized ($<1 \%$) at 22 GHz. 

The fractional polarization ratios generally lie between 0.8 and 1.8,
with a median value of $\sim 1.2$. We find these ratios are related to
the spatial resolution of our images, with the low-redshift sources
displaying significantly higher values (Fig.~\ref{depol_vs_angscale}).
\cite{T91} has shown that polarization fluctuations can occur in
uniform (non-clumpy) external Faraday screens, due to turbulent
magnetic fields. It is possible that in higher-redshift sources in our
sample, these fluctuations are being smeared out by the larger spatial
extent of the beam, leading to lower average depolarization
factors. This effect needs to be investigated further using larger
samples, as multi-frequency polarimetry is potentially a very useful
tool for probing the material in the central regions of AGNs.

\section{Observed differences between high- and low-optically
polarized quasars\label{HPQvsLPRQ}}

Various claims have been made in the literature that the differences
in high- and low-optically polarized radio quasars are due to their jets
having different viewing angles, with LPRQs lying farther
away from the line of sight. Supporting evidence for this scenario
includes differences in their variability timescales \citep{VTU92},
and their distributions of jet misalignments between parsec and
kiloparsec scales \citep{ILT91,XU94}. In this section we examine this
hypothesis, and show that it cannot fully account for differences seen
in the parsec-scale magnetic fields of LPRQs and HPQs.

\subsection{Core properties}

In section~\ref{coreopt} we showed that the level of optical
polarization in compact quasars is closely related to that of the
radio core at 43 GHz. The correlation in Fig.~\ref{core_m_vs_m_opt}
suggests that the optically polarized and radio emission arise from
the same process, (i.e., synchrotron radiation), and may in fact be
co-spatial. They must also be beamed by a similar amount, since any
difference in Doppler factor would change the observed fractional
polarization (due to aberration) and destroy the correlation. A likely
scenario, presented in previous studies
\citep[e.g.,][]{GSS96}, 
is that both the radio and optically-polarized flux arise from a
single energy distribution of electrons, located very close to the
base of the jet. The rapid variability and high degrees of optical
polarization seen in most blazars imply that the polarized emission
arises in a very small region having a well-ordered magnetic
field. These conditions are most readily created in a relativistic
shock, in which the level of observed polarization depends on both the
shock strength and viewing angle in the source frame
\citep[e.g.,][]{HAA85}.

The continuous trend in Figure~\ref{core_m_vs_m_opt} suggests
that the division between HPQs and LPRQs at $m_{\rm opt} = 3\%$ may be
an arbitrary one in the case of compact radio quasars.
Nevertheless, we will retain this classification criterion for the
purposes of comparing the general properties of high- and
low-polarization quasars in this paper.

We have performed Kolmogorov-Smirnov tests on various properties of
our HPQ and LPRQ samples to determine the probability that
they are drawn from the same population. We list the results of these
tests in Table~\ref{ks_results}.  The cores of our LPRQ sample in our
sample are fainter on average, and are more weakly polarized than
those of the HPQs. As a result, the LPRQs are less core-dominated, and
have steeper total spectral indices between 22 and 43 GHz. Although it
appears that the core magnetic fields of HPQs favor perpendicular
orientations with respect to the jet (see Fig.~\ref{dpa_core}), there
are insufficient data to characterize those of the LPRQs.

\begin{figure*}
\epsscale{1.0}
\plotone{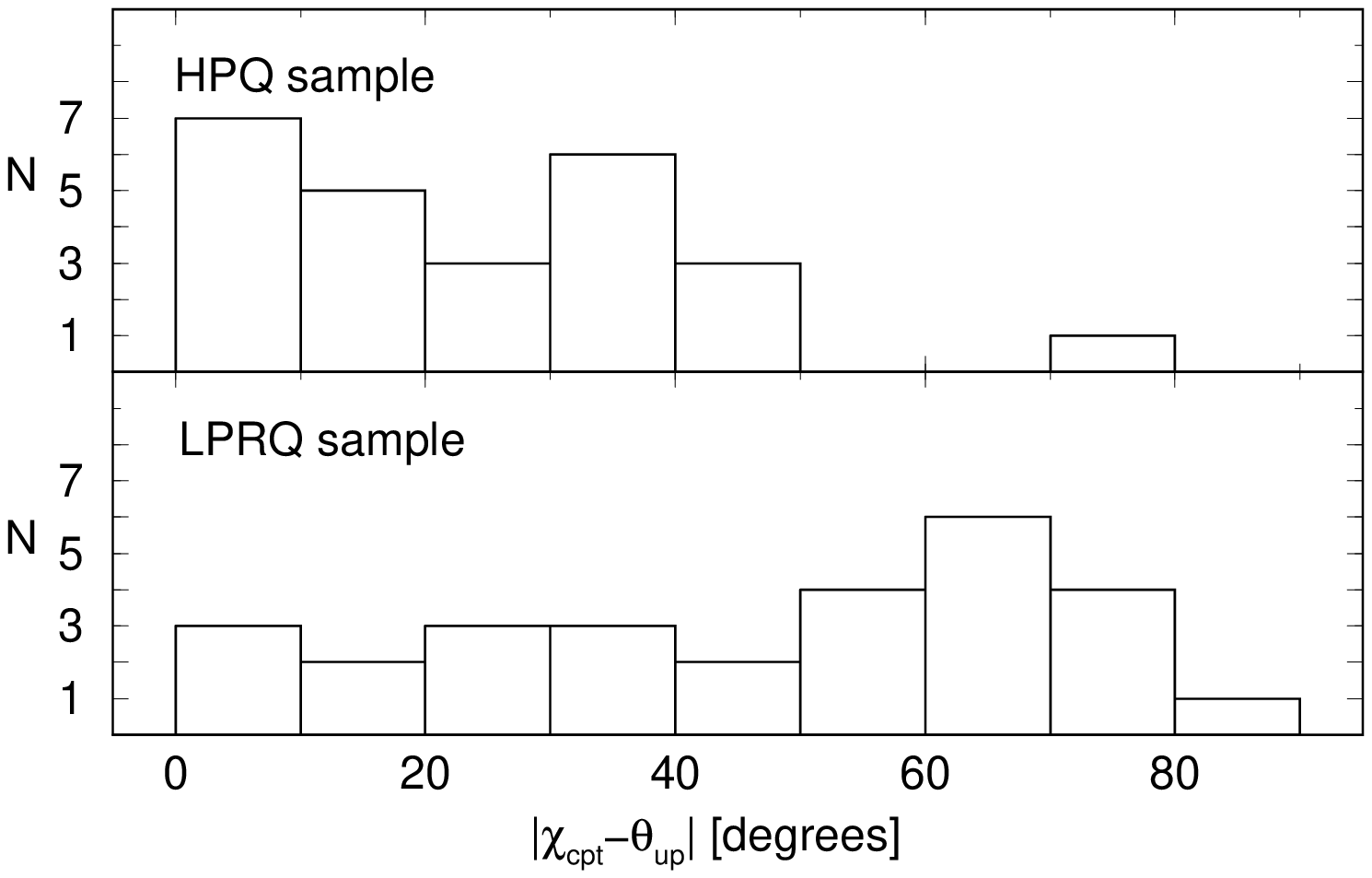}
\caption{\label{dPA43_up} Distribution of 43 GHz electric polarization vector
misalignment of polarized jet components with respect to the
local upstream jet direction ($\theta_{\rm up}$) for HPQs (top) and
LPRQs (bottom).}
\end{figure*}

\begin{figure*}
\epsscale{1.0}
\plotone{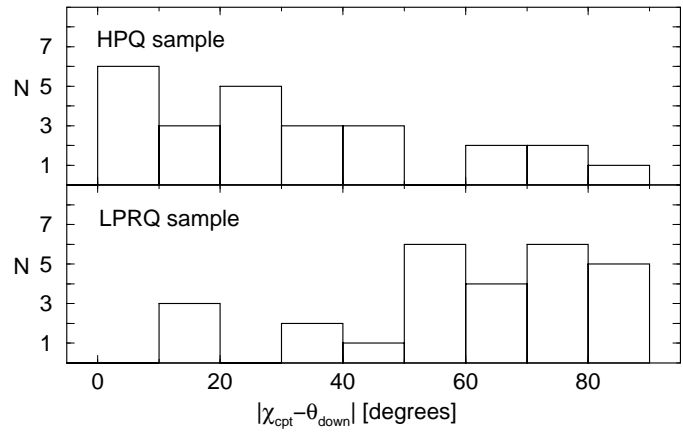}
\caption{\label{dPA43_down}Same as
Fig.~\ref{dPA43_up}, using the local downstream jet direction $\theta_{\rm down}$.}
\end{figure*}

\subsection{Jet Bending\label{bending}}
Many bright, compact radio quasars are known to exhibit large apparent
bends in their jets that are likely exaggerated by projection effects
associated with small angles to the line of sight. The amount of
apparent jet bending in a sample of objects should therefore be
statistically related to the mean viewing angle. Previous bending
studies
\citep[e.g.][]{TME98} have used two main statistics to
quantify the amount of bending: $\phi_T$, which represents the total
amount of bending along the jet (in degrees), and $\Delta\theta$,
which is defined as the difference in the innermost jet position angle
on parsec and kiloparsec scales.  Neither of these parameters is
wholly satisfactory, since they may be influenced by the spatial
resolution and dynamic range of the radio images used to calculate
them. In order to lessen the effects of the latter, in this paper we
measure $\phi_T$ out to 15 (projected) parsecs along the jet. We use
the best images available in the literature to measure the jet
position angles on kiloparsec scales. We list these data in
Table~\ref{totaldata}.

Despite the potential shortcomings of these bending parameters, we
find them to be correlated with several other source properties. The
more misaligned jets in our sample tend to have larger
core-to-extended flux ratios at 43 GHz, and flatter integrated 22/43
GHz spectral indices. They are also more polarized in the optical
regime (Fig.~\ref{bend_corr}). A KS test shows that the misalignment
distributions of our LPRQs and HPQs differ at the 98.7\% confidence
level.  A similar result was found by \citet{ILT91} for the
Pearson-Readhead sample, and \citet{XU94} for the much larger
Caltech-Jodrell survey.

To first order, these trends are easily explained by the orientation
model: HPQs, having smaller viewing angles, should have larger apparent
pc-kpc jet misalignments, and more highly beamed cores. This
will also increase their core-to-extended flux ratios, and flatten their
overall spectral indices. In the optical, the polarized flux from the
core would be boosted with respect to the continuum from the accretion
disk, which would raise the overall level of polarization. Upon deeper
inspection, however, this model displays several shortcomings.  If the
polarized flux in the core arises from one or more planar shocks, the
inner jets of LPRQs, lying farther away from the line of sight than the
HPQs, should have {\it higher} fractional polarizations, according to
the standard shock model of \citet{HAA85}. This is the opposite of
what is observed. We also find no differences in the total amount of
bending out to 15 pc in LPRQs and HPQs, which is difficult to
reconcile with the orientation hypothesis. Finally, there are two
trends with $\phi_T$ that are present only in the LPRQ data: the
amount of total jet bending increases with core luminosity and optical
polarization level.

\subsection{Jet properties}
\begin{figure*}
\epsscale{1.0}
\plotone{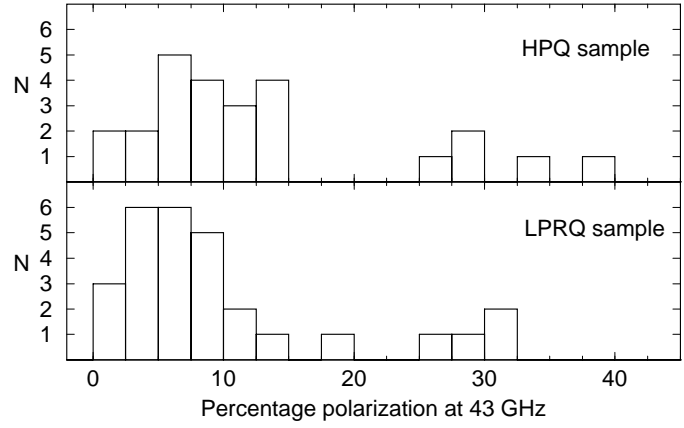}
\caption{\label{m43} Distribution of percentage
polarization for jet components in HPQs (top) and LPRQs (bottom).}
\end{figure*}

In addition to having different core properties, HPQs and LPRQs show
fundamental differences in their parsec-scale jets. First, the median
jet component luminosity of the LPRQs (26.2 W/Hz) is marginally
lower than the HPQs (26.7 W/Hz), and second, the polarized
components in HPQ jets tend to have magnetic fields that are
perpendicular to the jet (Figs.~\ref{dPA43_up} and
\ref{dPA43_down}). Those of LPRQs have a tendency to be more parallel
to the jet, although this depends somewhat on whether we compare the
$\chi$'s to the upstream or downstream jet direction. The percentage
polarizations of the jet components are similar for the two classes
(Fig.~\ref{m43}).

These last two findings effectively rule out the possibility that the
differences between LPRQs and HPQs are due to orientation alone. If
LPRQ and HPQ jets are intrinsically identical, and seen at different
viewing angles, we expect to see large differences in observed
fractional polarization, since this quantity varies as
$\cos^2{\epsilon} /( A - \cos^2{\epsilon})$, where ${\epsilon}$ is the
angle between the plane of the shock and the line of sight in the rest
frame of the emitting gas, and $A$ is a function of shock strength
\citep{HAA85}.

\begin{figure*}
\epsscale{1.0}
\plotone{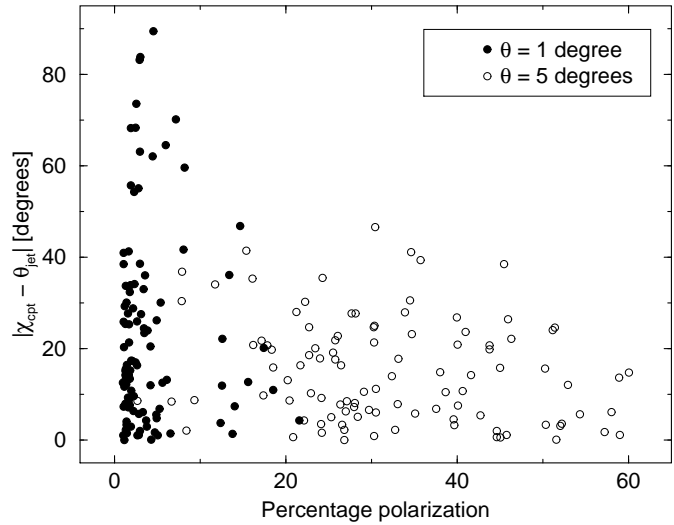}
\caption{\label{shockmodel} Simulated electric polarization offset with
respect to local jet direction, plotted against fractional
polarization for a population of moving, oblique shocks. The filled
circles represent a viewing angle of 1 degree to the jet, while the
empty circles represent the same jet viewed at 5 degrees from the line
of sight.}
\end{figure*}

It is also not possible to transform the observed
$|\chi_{\rm cpt}-\theta_{jet}|$ distribution for HPQs into that of the
LPRQs simply by changing the jet orientation. Indeed, in the case of
strong transverse shocks, the observed electric vectors will remain
parallel to the jet, independent of viewing angle \citep{BK79}.  To
produce a change in $\chi$ with orientation it is necessary to invoke
either an underlying longitudinal magnetic field
\citep[e.g.,][]{KWR90}, or oblique shocks. We have used the model of
\cite{LMG98} to calculate the observed $\chi$'s and percentage
polarizations of a population of moving, oblique shocks at two
different viewing angles. These shocks move at a variety of speeds
down a jet whose underlying flow has a bulk Lorentz factor of 15. In
Figure~\ref{shockmodel} we show a plot of $|\chi -
\theta|$ vs. percentage polarization for the same family of shocks,
seen at 1 degree from the line of sight (filled circles) and 5 degrees
(open circles). As the viewing angle is increased, the components
become more polarized, and their $\chi$'s become more aligned with the
jet. This behavior is inconsistent with a model in which LPRQs have
larger viewing angles than their HPQ counterparts.

\subsection{Discussion}

Many of the properties of HPQs and LPRQs that we have discussed in this
paper appear difficult to reconcile using a single model. While the
bending properties and core luminosities suggest possible differences
in viewing angle and Doppler factor, the jet polarization properties
suggest otherwise. In this section we discuss possible ways in which
to reconcile these seemingly contradictory findings.

\subsubsection{Shock strengths}
The simplest way to account for the jet polarization properties of
LPRQs is to assume that their shocks are weaker than those of HPQs, and
increase in strength with distance from the core. Their underlying jet
magnetic field cannot be randomly oriented, but must be somewhat
longitudinal to account for the $\chi$ orientations near the core. This
scenario can explain the lower core polarizations and luminosities in
LPRQs, and reproduces the trends of polarization and $\chi$ alignment
with distance down the jet. The differences in shock strengths may
also be responsible for the different optical $\chi$ misalignment
distributions of LPRQs and HPQs reported by \citet{ILT91}. A recent
high-resolution optical and radio study of M87's jet by \citet{PBZ99}
showed that the optically polarized emission in the inner jet comes
primarily from bright, shocked regions with magnetic fields
perpendicular to the jet. Strong shocks are apparently required to
boost and maintain the electron energies at levels where they are able
to emit optical synchrotron radiation.

It is also possible to incorporate this shock strength model into the
evolutionary scheme of \citet{F88}, who found that at any given epoch,
only $\sim 2/3$ of flat-spectrum quasars show blazar-like
properties. If the LPRQs in our sample are in a temporary quiescent
state, the more polarized shocks located down the jet may be relics
from previous ``blazar'' epochs when the source was emitting more
strongly-shocked components. We would also expect to see LPRQs that
have recently transformed into HPQs by emitting shocked components:
the former LPRQ 1633+382 may be an example of such an object. Our
images reveal a highly polarized component near the core that may have
triggered its dramatic increase in optical polarization.  The ``duty
cycle'' (the fraction of time that $m_{\rm opt}$ is $> 3\%$) of LPRQs
is still not well determined, however, as there have been no
systematic, long-term polarimetric monitoring studies of these
objects. Polarization variability is likely to be a better indicator
of blazar activity than the traditional $m_{\rm opt} > 3\%$ criterion,
since the latter can be affected by the strength of the
non-synchrotron optical continuum.

\subsubsection{Jet orientation}
Although we cannot completely rule out possible differences in the
orientation of HPQ and LPRQ jets, we consider it unlikely that they
differ by a large amount. Any change in viewing angle would have to be
offset by a complimentary change in shock or longitudinal field
strength, or alternatively, jet speed, in order to reproduce the
similar radio component polarization levels seen in both classes. The
similarities in redshift, emission-line equivalent width, and optical
luminosities of HPQs and LPRQs \citep{MS84,WWB92} suggest that they
undergo similar amounts of relativistic beaming. Also, \citet{MBP93}
find no differences in their core-to-extended flux ratios, which should
be a good indicator of beaming \citep[e.g.,][]{AU85}.

One possibility is that the observed differences in the
parsec/kiloparsec-scale jet misalignment angle could simply be due to
a higher degree of intrinsic bending in HPQs, and not
orientation. However, given the similarities in the $\phi_{T}$
distributions, this does not seem likely. Alternatively, the LPRQs may
represent jets whose bends lie in a plane perpendicular to that of the
sky. Monte Carlo simulations of bent jet populations \citep{L99} have
shown that flux-limited samples should contain a certain fraction of
these AGNs, whose innermost jets start out with moderately high
viewing angles, but then bend into the line of sight. In these cases,
the highly-aligned portion of the jet would likely obscure any
upstream emission, and might be mistakenly identified as the core due
to its large Doppler factor. The parsec-scale jets of these sources
would have the same small viewing angles as typical HPQs, but would
fade out faster with core distance, and be much better aligned with
their kiloparsec-scale structure. We are currently gathering 43 GHz
VLBI polarimetry and space-VLBI data on a large, complete sample of
LPRQs and HPQs (the Pearson-Readhead survey; \citealt{PR88}) with
which we will be able to investigate these scenarios more fully.

\section{Conclusions}
In this paper we have used high-frequency VLBI and optical polarimetry
to compare the parsec-scale magnetic field properties of a sample of
high-optical polarization, compact radio quasars with a sample of
similar objects having low optical polarizations. We summarize our
results as follows:

1. We find a strong correlation between the level of optical
polarization and radio core polarization at 43 GHz. The more optically
polarized quasars also have higher core luminosities, core-to-extended flux
ratios, and flatter integrated 22/43 GHz spectral indices. These
trends strongly indicate that the optically polarized emission is
synchrotron radiation, and is co-spatial with the radio core emission
at 43 GHz. 

2. The electric vectors of the highly polarized 43 GHz radio cores are
roughly aligned with the inner jet, indicating magnetic fields
perpendicular to the flow. A similar configuration is seen in the
optical, suggesting that the polarized flux at both wavelengths is due
to one or more strong transverse shocks located very close to the base
of the jet.

3. There is a strong trend for the fractional polarizations of bright
jet components to increase downstream from the core. The fact that all
of the components with magnetic fields parallel to the jet are located
near the core suggests that this trend is not due to an underlying
longitudinal magnetic field of increasing strength. Rather, we believe
it is either the result of the jet curving away from the line of
sight, or an increase in shock strength along the jet.

4. We find evidence for large rotation measures (up to $\sim 3000 \
\rm rad \ m^{-2}$) in the nuclear regions of low-optically polarized
radio quasars, which are indicative of parsec-scale Faraday screens
with organized magnetic fields.  The low-redshift quasars in our
sample tend to have jet components with larger 43/22 GHz
depolarization ratios than those found in the high-redshift sources.
This might be due to small-scale magnetic field fluctuations in the
Faraday screens that are being smeared out in the high-redshift
sources by the poorer spatial resolution of the restoring beam.

5. We find that the parsec-scale jet properties of compact radio
quasars are highly dependent on their level of optical
polarization. Sources with optical fractional polarizations below
$3\%$ (low-optically polarized radio quasars: LPRQs) tend to have
lower 43 GHz core polarizations, fainter cores, steeper total spectral
indices, smaller pc/kpc-scale jet misalignments, and smaller
core-to-extended flux ratios than high-optically polarized quasars
(HPQs). Although the components in the jets of HPQs and LPRQs have
similar fractional polarizations, those found in HPQs tend to have
magnetic fields that are perpendicular to the jet, while those in LPRQ
jets have mainly parallel orientations.

6. The observed differences in LPRQs and HPQs cannot be fully
explained by a model in which LPRQs are seen at larger angles from the
line of sight. Instead, our data are more consistent with an
evolutionary scenario based on \citet{F88}, in which flat-spectrum
quasars go through episodic stages of blazar-like activity. During
these phases, they emit strongly shocked components with transverse
magnetic fields, which move at a variety of speeds down the jet. LPRQs
may represent quiescent phases of blazars in which only weak shocks
are generated in the flow.

\acknowledgments

 We wish to thank Antxon Alberdi for sharing polarization data on
 4C~39.25 prior to publication.  We also wish to thank Harri
 Ter\"asranta of the Mets\"ahovi Observatory for kindly providing
 single dish flux densities at 22 and 37 GHz for many of our sample
 sources.

 We thank the Director of Steward Observatory for access to the 60" telescope,
 and Jim Grantham and Bob Peterson for the maintenance of this facility.
 We also thank Gary Schmidt for maintaining and for allowing us to use
 the Two-Holer polarimeter.  PSS acknowledges partial support of this
 research from a Lucas Junior Faculty Award.

 This research was performed in part at the Jet Propulsion Laboratory,
 California Institute of Technology, under contract to NASA, and has
 made use of data from the following sources:

 The NASA/IPAC Extragalactic Database (NED), which is operated by the
 Jet Propulsion Laboratory, California Institute of Technology, under
 contract with the National Aeronautics and Space Administration.

 The University of Michigan Radio Astronomy Observatory, which
 is supported by the National Science Foundation and by funds from the
 University of Michigan.


\begin{deluxetable}{llrrrrrrrrr}
\tablecolumns{11}
\tablecaption{\label{totaldata}Overall Radio Source Properties}
\tablewidth{0pt}
\tablehead{ \colhead{Source} & \colhead{Epoch} & \colhead{z} &
\colhead{$\alpha$} &  \colhead{ $L_{22}$}&\colhead{ $L_{43}$} 
&\colhead{$m_{43}$} & \colhead{ $\theta_{kpc}$} & \colhead{
$\theta_{pc}$} &  \colhead{ $\phi_{tot}$} & \colhead{ Ref.}  \\
\colhead{ [1]} & \colhead{ [2]} & \colhead{ [3]}  &
\colhead{ [4]} & \colhead{ [5]} & \colhead{ [6]} &
\colhead{ [7]}  & \colhead{ [8]}  & \colhead{ [9]}&\colhead{ [10]} & \colhead{[11]} }

\startdata
\sidehead{Low polarization quasars}
NRAO 140 & 1999 Jan 12 & 1.258 & $-0.36$ &27.84&27.71  & 0.8& 143  &
138 & 28 & 1 \\
4C 39.25 &1999 Jan 12 & 0.695 & 0.82    &28.10&27.93   & 7.1&
76  & 68 & 40 & 2 \\ 
0953+254 & 1999 Jan 12 & 0.712 & 0.43    &27.15&27.07      &11.9&$-109$&
$-124$  & 68 & 1 \\
3C 273   &1999 Jan 12& 0.158 &$-0.11$  &27.40&27.24          & 6.6&$-130$&
$-128$ & 15& 3  \\
1611+343 &1999 Jan 12 & 1.401 &$-0.12$  &28.27&28.11          &
9.0&$-170$& 169 & 170& 1 \\
1928+738 & 1999 Jan 12 &0.302 &$-0.07$  &26.87&26.74      &
2.8&$-170$&  166 & 177& 4 \\
2134+004 &1999 Jan 12 &  1.932 &0.83     &28.74&28.59     & 7.4&\n
&$-29$ &87 &\n\\
2145+067 & 1999 Jan 12 & 0.990 &0.29     &28.38&28.38     & 4.4&\n
&   72 &70 &\n  \\
2201+315 & 1999 Jan 12& 0.295 &0.20     &26.82&26.70     &
3.7&$-127$&$-153$ &14 &1\\
\sidehead{High polarization quasars}

0420--014 &1998 Jul 31 &0.915 &0.67     &27.96&27.88         & 1.7&
171&$-132$ & 55 &6 \\
1055+018  &1996 Nov 23&0.888 &0.22     &\n&27.83    & 9.0&
$-179$ & $-63$ &  20 &1  \\
3C 279    &1999 Jan 12 &0.536 &0.02     &28.31&28.26         &11.5&
$-153$&$-128$ & 27 &5 \\ 
1334--127 &1996 Nov 23 &0.539 &0.13     &\n&27.90    & 5.2& \n
& 139 & 12 & \n  \\
1510--089 &1998 Jul 31& 0.360 &$-0.01$  &26.74&26.75           & 4.0&
163 & $-38$ &  57 &7 \\
1633+382  &1999 Jan 12 &1.814 &0.42     &28.27&28.29         & 7.0&
176&$-81$ & 136 &1 \\
3C 345    &1998 Jul 31 &0.593 &0.05     &27.89&27.86        & 5.3& $-32$ &
$ -76$ &163 &8 \\
CTA 102   &1998 Jul 31 &1.037 &$-0.46$  &28.18&28.15      & 4.3&
147 & 108 & 41 &6 \\
3C 454.3   &1998 Jul 31 &0.859 &0.11     &28.31&28.23    & 6.4&
$-61$ & $-80$& 35&1 \\  
\enddata

\tablecomments{Luminosities are calculated assuming $H_o =
65\ \rm km \ s^{-1} \ Mpc^{-1}$, $\Lambda = 0$, and $q_o = 0.1$. Columns are as
follows: (1) Source name. (2) UT date of VLBA radio observations.  (3) Redshift. (4) Spectral index between 1.4
and 5 GHz, obtained from NED. (5) Total luminosity at 22 GHz in $\rm W
\ Hz^{-1}$, from Mets\"ahovi single dish data.  (6) Parsec-scale luminosity at 43
GHz in $\rm W
\ Hz^{-1}$, derived from cleaned flux in VLBA maps.  (7) Parsec-scale
integrated percentage polarization at 43 GHz. (8) Innermost jet
position angle on kiloparsec scale. (9) Innermost jet position angle
on parsec scale. (10) Total amount of bending along jet (in degrees)
out to a projected distance of 15 pc from the core. (11) Reference for
image used to measure values in columns 8 and 10. }

\tablerefs{(1) \citealt{MBP93}; (2) \citealt{MZS91}; (3) \citealt{DMC85};
(4) \citealt*{KWR90}; (5) \citealt{DP83}; (6) \citealt{AU85}; (7)
\citealt{PGH93}; (8) \citealt*{KWR89}.}
\end{deluxetable}
\begin{deluxetable}{lllrrrrrrrl}
\tabletypesize{\scriptsize}
\rotate
\tablecolumns{11}
\tablecaption{\label{Kmapdata}Summary of 22 GHz VLBA image parameters}
\tablewidth{0pt}
\tablehead{ 
\colhead{ Source} & \colhead{ Type} &
\colhead{ Weight} &
\colhead{ Beam}&  \colhead{ PA} &
\colhead{ Flux} & \colhead{ EV} & \colhead{ RMS} &
\colhead{ DNR} & \colhead{ Peak} &
\colhead{ Contour levels}\\
\colhead{ [1]} & \colhead{ [2]} & \colhead{ [3]}  &
\colhead{ [4]} & \colhead{ [5]} &
\colhead{ [6]} & \colhead{ [7]}  & 
\colhead{ [8]} & \colhead{ [9]} & \colhead{ [10]}
 & \colhead{ [11]}}

\startdata

  NRAO~140&   IPOL &  Natural
&   0.85 x 0.43 &  $  -20$&   1.594
&      &   0.4 &   3559& 
1238 &   $-0.1$, 0.1, 0.2, 0.4, 0.8,
1.6, 3.2, 6.4, 12.8, 95\\
        &   PPOL &      &               &             &   0.029 &  17  &   0.3 &   24  &   7.1\\
  4C 39.25&   IPOL &  Uniform&   0.60 x 0.35 &  $  -10$ &  10.415 &      & 
2.0 &   2525&   5150 & $-0.15$, 0.15, 0.3, 0.6, 1.2, 2.4, 4.8, 9.6,
19.2, 38.4, 76.8 \\
        &   PPOL &      &               &             &   0.374 &  267 &   1.2 &   180 &    215 \\
  0953+254&   IPOL &  Uniform&   0.83 x 0.35 &  $  -22$ &   1.078 &      & 
0.8 &   818 &   648 &  $-0.3$, 0.3, 0.6, 1.2, 2.4,
4.8, 9.6, 19.2, 38.4, 76.8 \\
        &   PPOL &      &               &             &   0.038 &  100 &   0.6 &   26&   15.6\\
  3C 273  &   IPOL &  Uniform&   0.97 x 0.38 &  $  -16$ &  35.237 &      & 
6.1 &   2820 &   17256 & $-0.15$, 0.15, 0.3, 0.6, 1.2, 2.4, 4.8, 9.6, 19.2,
38.4, 76.8 \\
        &   PPOL &      &               &             &   1.699 &  400 &   3.4 &   217 &   738 \\
  3C 279  &   IPOL &  Uniform&   0.99 x 0.35 &  $ -11 $ &  27.649 &      & 
3.6 &   4108 &   14789 & $-0.1$, 0.1, 0.2, 0.4, 0.8,
1.6, 3.2, 6.4, 12.8, 25.6, 51.2 \\
        &   PPOL &      &               &             &   2.342 &  2000&   3.8 &   278 &   1056\\
  1611+343&   IPOL &  Uniform&   0.64 x 0.34 &  $ -30 $ &   3.764 &      & 
1.1 &   2067 &   2357 & $-0.2$, 0.2, 0.4, 0.8,
1.6, 3.2, 6.4, 12.8, 25.6, 51.2 \\
        &   PPOL &      &               &             &   0.156 &  133 &   0.8 &   51 &   41\\
 1633+382&   IPOL &  Uniform&   0.54 x 0.34 &  $ -29 $ &   2.194 &      & 
0.7 &   1785 &   1391 & $-0.15$, 0.15, 0.3, 0.6, 1.2,
2.4, 4.8, 9.6, 19.2, 38.4, 76.8 \\
        &   PPOL &      &               &             &   0.054 &  40  &   0.5 &   48 &   24\\
 1928+738&   IPOL &  Natural &   0.56 x 0.49 &  $ -88 $&   3.196 &      & 
0.8 &   2209 &   1747 & $-0.25$, 0.25, 0.5, 1, 2,4,
8, 16, 32, 64 \\
        &   PPOL &      &               &             &   0.061 &  67  &   0.4 &   30 &   12\\
 2134+004&   IPOL &  Uniform&   1.36 x 0.32 &  $ -20 $ &   5.503 &      & 
1.6 &   1241 &   1936 &  $-0.45$, 0.45, 0.9, 1.8, 3.6,
7.2, 14.4, 28.8, 57.6\\
        &   PPOL &      &               &             &   0.314 &  400 &   0.8 &   174 &   139\\
 2145+067&   IPOL &  Uniform&   1.11 x 0.34 &  $ -21 $ &   9.245 &      & 
1.4 &   4538 &   6551 &  $-0.1$, 0.1, 0.2, 0.4, 0.8,
1.6, 3.2, 6.4, 12.8, 25.6, 51.2\\
        &   PPOL &      &               &             &   0.265 &  500 &   1.2 &   120 &   144\\
 2201+315&   IPOL &  Uniform&   0.69 x 0.38 &  $ -35 $ &   2.872 &      & 
0.8 &   1181 &   1032 & $-0.35$, 0.35, 0.7, 1.4, 2.8,
5.6, 11.2, 22.4, 44.8, 89.6 \\
        &   PPOL &      &               &             &   0.077 &  100 &   0.6 &   82 &   49\\
\enddata
\tablecomments{Columns are as follows: (1) Source
name. (2) Image polarization type. (3) Visibility weighting scheme of
image. (4) FWHM dimensions of Gaussian
restoring beam, in mas. (5) Position angle of restoring beam, in
degrees. (6) Total cleaned flux density in Janskys. (7) Electric
vector scaling in image [$\rm mJy \ mas^{-1}$]. (8) RMS noise level
[$\rm mJy \ beam^{-1}$].  (9) Dynamic range
(peak/RMS). (10) Peak intensity [$\rm mJy \ beam^{-1}$]. (11) Contour levels,
expressed as a percentage of peak intensity. }

\end{deluxetable}


\begin{deluxetable}{lllrrrrrrrl}
\tabletypesize{\scriptsize}
\rotate
\tablecolumns{11}
\tablecaption{\label{Qmapdata}Summary of 43 GHz VLBA image parameters}
\tablewidth{0pt}
\tablehead{
\colhead{\ Source} & \colhead{\ Type} &
\colhead{\ Weight} &
\colhead{\ Beam}&  \colhead{\ PA} &
\colhead{\ Flux} & \colhead{\ EV} & \colhead{\ RMS} &
\colhead{\ DNR} & \colhead{\ Peak} &
\colhead{\ Contour levels}\\
\colhead{\ [1]} & \colhead{\ [2]} & \colhead{\ [3]}  &
\colhead{\ [4]} & \colhead{\ [5]} &
\colhead{\ [6]} & \colhead{\ [7]}  & 
\colhead{\ [8]} & \colhead{\ [9]} & \colhead{\ [10]}
 & \colhead{\ [11]}}

\startdata
\ NRAO 140&\  IPOL &\ Natural &\  0.35 x 0.25 &\ $ -21$ &\  1.259  &\ &\  1.0 &\   843 &\ 792&\ $-0.35$,
0.35, 0.7, 1.4, 2.8, 5.6, 11.2, 22.4, 44.8, 89.6\\
        &\  PPOL &\     &\              &\             &\ 0.011 &\ 117 &\  0.6 &\ 
14 &\  8.6&\ \\
\ 4C 39.25&\  IPOL &\ Uniform&\  0.29 x 0.17 &\ $ -1$   &\  6.986  &\ &\  1.3 &\  1049&\  2204 &\ $-0.4$, 0.4, 0.8, 1.6, 3.2,
6.4,12.8, 25.6,  51.2\\
        &\  PPOL &\     &\              &\           &\  0.394  &\ 233 &\  1.5 &\ 
107&\  160 &\ \\
\ 0953+254&\  IPOL &\ Natural &\  0.38 x 0.22 &\ $  -7 $&\  0.925  &\ &\  0.7 &\   842 &\ 589&\ $-0.45$, 0.45, 0.9, 1.8, 3.6, 7.2, 14.4, 28.8, 
57.6\\
        &\  PPOL &\     &\              &\            &\  0.012 &\ 78 &\  0.7 &\    15
&\  10.4&\ \\
\ 3C 273  &\  IPOL &\ Natural &\  0.52 x 0.21 &\ $ -11 $&\ 27.283  &\ &\  8.1 &\  1063 &\ 
8635&\ $-0.35$, 0.35, 0.7, 1.4, 2.8,
5.6,11.2, 22.4, 44.8,  89.6\\
        &\  PPOL &\     &\              &\            &\ 1.576 &\  933 &\  5.0 &\ 
93&\  465&\ \\\
\ 3C 279  &\  IPOL &\ Natural &\  0.47 x 0.18 &\ $ -9  $&\ 25.255 &\  &\  3.2 &\  4300 &\ 13413&\ $-0.125$, 0.125, 0.25, 0.5, 1, 2,4, 8, 16,  95\\
        &\  PPOL &\     &\              &\            &\ 2.681 &\  1944 &\  5.7 &\ 
154 &\  879&\ \\
\ 1611+343&\  IPOL &\ Natural &\  0.36 x 0.21 &\ $ -9  $&\  2.577 &\  &\  1.0 &\  1557 &\ 1504&\ $-0.2$, 0.2, 0.4, 0.8,
1.6, 3.2, 6.4, 12.8, 25.6,  51.2\\
        &\  PPOL &\     &\              &\            &\ 0.182 &\  117 &\  0.9 &\    69 &\ 62&\ \\
\ 1633+382&\  IPOL &\ Natural &\  0.33 x 0.21 &\ $ -13 $&\  2.286 &\  &\  0.7 &\  2340 &\ 1736&\ $-0.125$, 0.125, 0.25, 0.5, 1, 2,4, 8, 16,  95\\
        &\  PPOL &\     &\              &\            &\ 0.067 &\  156 &\  0.9 &\    48 &\ 43&\ \\
\ 1928+738&\  IPOL &\ Natural &\  0.27 x 0.25 &\ $ 54  $&\  2.423 &\  &\  0.8 &\  1503 &\ 1284&\ $-0.25$, 0.25, 0.5, 1, 2,4,
8, 16, 32,  64\\
        &\  PPOL &\     &\              &\            &\ 0.076 &\  156 &\  0.8 &\    46&\ 37 &\ \\
\ 2134+004&\  IPOL &\ Natural &\  0.47 x 0.19 &\ $  -8 $&\  3.980 &\  &\  1.0 &\  1284&\ 1348 &\ $-0.3$, 0.3, 0.6, 1.2, 2.4,
4.8, 9.6, 19.2, 38.4,  76.8\\
        &\  PPOL &\     &\              &\            &\ 0.247&\  156  &\  0.9 &\    83&\ 75 &\ \\
\ 2145+067&\  IPOL &\ Uniform&\  0.38 x 0.17 &\ $  -9 $ &\  9.636  &\ &\  2.7 &\  2264 &\ 6021&\ $-0.2$, 0.2, 0.4, 0.8,
1.6, 3.2, 6.4, 12.8, 25.6,  51.2\\
        &\  PPOL &\     &\              &\            &\ 0.365 &\  583 &\  2.0 &\   118 &\ 236&\ \\
\ 2201+315&\  IPOL &\ Natural &\  0.38 x 0.20 &\ $ -5  $&\  2.275  &\ &\  0.7 &\  1121 &\ 816&\ $-0.3$, 0.3, 0.6, 1.2, 2.4,
4.8, 9.6, 19.2, 38.4,  76.8\\
        &\  PPOL &\     &\              &\            &\ 0.065 &\  233 &\  0.7 &\    54 &\ 38&\ \\
\enddata

\tablecomments{Columns are as follows: (1) Source
name. (2) Image polarization type. (3) Visibility weighting scheme of
image. (4) FWHM dimensions of Gaussian
restoring beam, in mas. (5) Position angle of restoring beam, in
degrees. (6) Total cleaned flux density in Janskys. (7) Electric
vector scaling in image [$\rm mJy \ mas^{-1}$]. (8) RMS noise level
[$\rm mJy \ beam^{-1}$].  (9) Dynamic range
(peak/RMS). (10) Peak intensity [$\rm mJy \ beam^{-1}$]. (11) Contour levels,
expressed as a percentage of peak intensity. }

\end{deluxetable}


\begin{deluxetable}{lrrrrlrrrrrrrrl}
\rotate
\tablewidth{0pt}

\tablecolumns{15}
\tablecaption{\label{coredata}Radio Core Component and Optical
Polarization Properties}
\tablehead{ \colhead{ Source} & \colhead{
$S_{22}$} &  \colhead{ $S_{43}$} & \colhead{
$L_{43}$} & \colhead{ $R_{43}$} &\colhead{ V} &\colhead{ $m_{22}$} &
\colhead{ $m_{43}$} & \colhead{ $m_{opt}$} &
\colhead{ $\chi_{22}$} &  \colhead{ $\chi_{43}$} &
\colhead{ $\chi_{opt}$} 
&\colhead{ $m_{43}/m_{22}$}
&\colhead{ $|\Delta\chi|$}&\colhead{ Ref.} \\
\colhead{ [1]} & \colhead{ [2]} & \colhead{ [3]}  &
\colhead{ [4]} & \colhead{ [5]} & \colhead{ [6]} & \colhead{ [7]}  &
\colhead{ [8]} & \colhead{ [9]} & \colhead{ [10]}
 & \colhead{ [11]} & \colhead{ [12]} & \colhead{ [13]}
 & \colhead{ [14]} & \colhead{  [15]}} 
\startdata
\sidehead{Low polarization quasars}
 NRAO 140 &  1358 &  123 & 26.70 & 0.11& 17.7&   \n & $<0.1$& $<0.9 \ (0.7)$
& \n & \n &  $148$ & \n& \n&  1\ (2) \\
 4C 39.25 &  209 &  298 & 26.56 & 0.04& 17.9&   $<0.6$& $<1.6$& $<0.8\ (0.4)$
& \n &  \n &  $116$ & \n& \n&  1\ (4) \\ 
 0953+254 & 686 &  634 & 26.91 & 2.18& 17.5&   2.4 &  1.7 &  1.4 & 
 3&  28  &  162  & 0.8  &  19&  1 \\
 3C 273   & 2792  & 2460  & 26.19  & 0.10& 12.7&  $<0.1$ &  $<0.6$&  0.3
 & \n& \n & \phn 56 & \n& \n&  1\\
 1611+343   & 2563  & 1686  & 27.93  & 1.89& 17.5&  1.0  &  3.8    &  2.3
  & $-20$ &  $-8$   &  165  & 2.0  &  8 &  1 \\
  1928+738   & 1966  & 298  & 25.83  & 0.14& $16.5$&  \n& $<0.8$& $0.8$  & \n& \n & $163$& \n& \n&  3\\
  2134+004   & 889  & 1329  & 28.11  & 0.50& $18.0$&  3.2 &  1.7    &  $2.8$ &  19   &   60
& \phn$86$&  0.5  &  11&  2 \\
  2145+067   & 6876  & 1890  & 27.67  & 0.24& $16.5$&  2.0 &  1.1    & 
$0.6$& 22&   30 &  $138$ & 1.7  &  2&  3 \\
  2201+315   & 586  & 818  & 26.25  & 0.56& $15.5$&  0.5 & $ <0.3$ &  $0.2$ &    53&    \n  & \phn
$80$ & \n& \n& 3  \\
\sidehead{High polarization quasars}

  0420--014 &   \n    &  2838 &  27.78 & 3.65& 18.0 & \n &  1.8 &   4.6 &  \n   & 
34  &  122  & \n& \n& 1\\
  1055+018  &    \n      &  1566 &  27.49 & 0.85& $18.0$& \n &       9.0 & 
$5.0$  &  \n   &  $-55$ &  $146$ & \n& \n& 3\\
  3C 279    &  14522 & 12069 &  27.94  & 0.92& 15.0&  3.9 &  6.6 &  36.9 
&  79&    82  &    75 & 1.7  &  2&  1\\ 
1334--127 &    \n      &  8381 &  27.79 & 3.22& $18.5$& \n &  5.0 &  $10.6$
 &  \n &   153  &  \phn \phn $8$ & \n& \n& 3 \\
  1510--089  &    \n      &  1434
&  26.67  & 4.82&  17.8& \n &  4.6 & 4.1  & \n &  3&  151 & \n& \n & 1
\\
  1633+382  &   1386 &  1754 &  28.18 & 3.16& 16.9&  1.7 &  2.5 &   7.0 & 38& 
$24$&  105  & 1.3  &  15& 1 \\
  3C 345     &   \n       &  6234 &  27.74  & 3.19& 16.8& \n &  5.0 & 11.6 &  \n   &    81  &  \phn 67& \n& \n& 1  \\
  CTA 102    &    \n      &  2307 &  27.80  & 0.79& $17.3$& \n &       2.9 & 
$7.3$ &  \n   &    66  &  $118$& \n& \n& 3 \\
  3C 454.3   &    \n      &  2346 &  27.64 & 0.34& $16.1$& \n &       1.6 & 
$2.9$  & \n &   $-71$&  $144$ & \n& \n& 3\\

\enddata

\tablecomments{Columns are as follows: (1) Source
name. (2-3) Core flux density at 22 and 43 GHz, in mJy. (4) Core
luminosity at 43 GHz [$\rm W \ Hz^{-1}$], (calculated assuming $\alpha = 0$). (5)
Ratio of core to extended flux density on parsec-scales at 43 GHz, in observer
frame. (6) Optical V magnitude.  (7-8) Percentage polarization of core
at 22 and 43 GHz. (9) Optical percentage polarization. (10-11) Electric
vector position angle of core at 22 and 43 GHz. (12) Optical electric
vector position angle. (13) Ratio of 43 and 22 GHz percentage
polarization, calculated using convolved 43 GHz image. (14) Difference
in electric vector position angle between 43 and 22 GHz, calculated
using convolved 43 GHz image. (15) Reference for optical data. }

\tablerefs{(1) This work; (2) \citet{WWE92}; (3) \citet{IT90}; (4) \citet{ILT91}.} 

\end{deluxetable}


\begin{deluxetable}{lrrrrrrrrrrrr}
\tablecolumns{13}
\tablewidth{0pt}
\tablecaption{\label{components}Jet Component Properties}
\tablehead{ \colhead{Cpt.} & \colhead {$r_{43}$} &
\colhead{$PA_{43}$} & \colhead{$S_{22}$} & \colhead{$S_{43}$} &
\colhead{$m_{22}$} & \colhead{$m_{43}$} & \colhead{$\chi_{22}$} &
\colhead{$\chi_{43}$} & \colhead{$\theta_{up}$} &\colhead{$\theta_{down}$} &
\colhead{$m_{43}/m_{22}$} & \colhead{$|\Delta\chi|$} \\
\colhead{(1)} & \colhead{(2)} & \colhead{(3)} & \colhead{(4)} &
\colhead{(5)} & \colhead{(6)} & \colhead{(7)} & \colhead{(8)} & 
\colhead{(9)} & \colhead{(10)} & \colhead{(11)} & \colhead{(12)}& \colhead{(13)}}
\startdata
\sidehead{NRAO140}
e& 0.27 &  139 &   \n&  799&   0.5&  0.8&  $-88$&   $-9$&   139 & 135&1.4 & 77 \\*
d& 0.53 &  135 &   \n&  274&   \n&   1.7&\n&  $ -7 $&   135 & 123&\n&\n\\*
C& 1.22 &  125 &   84&   32&  7.8&\n&    51&\n&   123 & 127&\n&\n\\*
B&2.84\tablenotemark{a}& 122\tablenotemark{a}&   58&   \n&  6.0&\n&   $ 23$&\n&   127 & 139&\n&\n\\*
A&5.27\tablenotemark{a}& 127\tablenotemark{a}&   26&   \n&   \n&\n&\n&\n&   133& 133&\n&\n\\
\sidehead{4C 39.25}
e& 1.05 &   93 &   \n&   36&   \n&\n&\n&\n&    93& 123&\n&\n\\*
d& 1.38 &  101 &   \n&   48&   \n&\n&\n&\n&   123& 135&\n&\n\\*
C& 1.79 &  109 &  138&  102&  9.9&\n&  $-10$&\n&    135 & 77&\n&\n\\*
b& 2.17 &  103 &   \n&   69&   \n&\n&\n&\n&    77&94&\n&\n\\*
A& 2.83 &  101 &4085&2027&  3.2&   5.7&    64&    71&    94& 81&     1.4&      4\\
\sidehead{0953+254}
c& 0.19 &$-124$&   \n&   46&   \n&   4.1&\n&  $-10$& $-124$&$-124$&\n&\n\\*
B& 0.52 &$-124$&  256&  148&  6.4&   3.4&  $-23$&  $-16$& $-124$&$-124$&     0.3&      1\\*
A& 1.13 &$-124$&   94&   76&   \n&\n&\n&\n&  $-124$&$-52$&\n&\n\\
\sidehead{3C 273}
j& 0.15 &$-128$&   \n& 1509&   \n&\n&\n&\n& $-128$&$-136$&\n&\n\\*
i& 0.32 &$-133$&   \n&  452&   \n&\n&\n&\n& $-136$&$-114$&\n&\n\\*
H& 0.80 &$-121$& 9282& 8340&  5.0&   9.4&  $-13$&  $-42$& $-114$& $-142$&     1.7&   $26$\\*
G& 1.16 &$-127$&19399&12492&  4.2&   6.2&  $-32$&  $-33$& $-142$& $-125$&    1.5&    $4$\\*
F& 1.40 &$-126$&  413& 1323&  3.3&   4.8&  $-43$&  $-59$&$-125$&$-125$ &   1.4&   $12$\\*
E& 1.90 &$-127$&  108&  118&  \n&\n&\n&\n& $-125$&$-125$&\n&\n\\*
D&3.78\tablenotemark{a}&$-119\tablenotemark{a}$&  226&   \n&  21.4&\n&  $-23$&\n& $-125$&$-125$&\n&\n\\*
C&6.44\tablenotemark{a}&$-112\tablenotemark{a}$&  276&   \n&  21.8&\n&  $-66$&\n& $-83$&$-112$&\n&\n\\*
B& 7.80 &$-112$ &  258&   95&   22.9&  32.1&  $-66$ & $-56$& $-112$&$-134$&  1.4&    $10$\\*
A&11.75\tablenotemark{a}&$-120\tablenotemark{a}$& 126&   \n&  13.9&\n&    $-63$ &\n& $-122$&$-122$&\n&\n\\
\sidehead{3C279}
G&  0.15&  $ -128$&   2526&   6035&   9.7&   9.3&    39&    61&  $-128$&$-131$  &  0.8&   10\\*
F&  0.64&  $-131$&   8083&   4991&   18.3&  25.7&    36&    35&  $-131$&$-131$  &  1.2&    1\\*
E&  0.89&  $ -120$&    185&    463&  12.7&  12.9&    50&    49&  $ -96$&$-168$  &  0.9&    $9$\\*
d&  1.15&  $-133$&\n&     70&\n&  34.1&\n&    51&  $-131$& $-109$&\n&\n\\*
C&1.66\tablenotemark{a}&  $-125\tablenotemark{a}$&     37&\n&   \n &\n&    \n&\n&  $-110$& $-110$&\n&\n\\*
B&3.34\tablenotemark{a}& $-118\tablenotemark{a}$&    464&    \n &  13.6&\n& $-82$&\n& $ -110$&$-110$&    1.2&   $2$\\*
A/C4&  3.73&   $-116$&   1840&   1418&  11.7&  13.9& $-86$& $-85$&
$-108$&$-108$ &   0.9&    $ 7$\\
\sidehead{1611+343}
F& 0.19 &  169 &  \n &  203&   \n &   0.5&    \n & $-32$&    169&96&   \n & \n \\*
E& 0.32 &  131 &  238&  232&  3.9&  14.2&$-82$ & $-88$& 96&173& 0.8 & $4$ \\*
D& 1.18 &  163 &   61&   27&\n&\n&\n&\n&   173&$-160$&\n&\n\\*
C& 2.93 &  172 &  175&  111&  20.0&  31.0&  $-22$&  $-31$&   169&160&     1.4&    $5$\\*
B& 3.97 &  169 &   68&   39&  17.8&  26.0&  $-29$&  $-32$&   160& 160&1.5&  $2$ \\*
A&4.14\tablenotemark{a}& 149\tablenotemark{a}&   42&   \n&\n&\n&\n&\n&   118&118&\n&\n\\
\sidehead{1633+382}
c2&  0.10&  $-81$&\n&     92&\n&   2.0&\n&   21   &   $-81$&$-96$&\n&\n\\*
c1&  0.47&    $-93$&    154&     41&   0.9&   8.7&    68& $-64$&  $ -96$&$-43$  &  3.1&   70\\*
B&  0.81&     $-69$&    102&     50&\n&\n&\n&\n&$-43$&$-92$&\n&\n\\*
A&  2.02&    $-91$&     92&     49&   4.0&\n&    52&\n&   $-92$&$-72$&\n&\n\\
\\
\\
\\
\\
\\

\sidehead{1928+738} 
G& 0.24 &  166 &   \n& 1338&   \n&   2.7&\n&  $-80$&   166&159&\n&\n\\*
f2&0.54 &  162 &   \n&  209&   \n&\n&\n&\n&   159&168&\n&\n\\*
f1&0.76 &  164 &  181&   14&   0.8&\n&   36  &\n&   168& 127&    1.4&     29\\*
E2&1.06\tablenotemark{a}& 153 &   \n&   87&   \n&   7.0&\n&    82&   127&159&\n&\n\\*
E& 1.42 &  155 &  583&  260&   0.6&\n&   7   &\n&   159& $-155$&  1.6&   $85$\\*
D& 1.66 &  164 &   56&   18&   3.0&  11.0  &   13  &  18   & $-155$&149  &   1.3&    9\\*
C& 2.77 &  158 &   90&   38&   7.4&\n&    85&\n&   149&$-106$& \n &\n\\*
B& 2.74 &  168 &   65&    8&  17.3&28.3&  70&71& $-106$&171&1.7 & 1 \\*
A& 3.20 &  168 &   89&   11&  15.5&  18.6&    49& $41$&   171&$-49$& 1.4&  $8$ \\
\sidehead{2134+004}
e& 0.24 & $-29$&   \n&  164&   \n&   3.2&\n&   46 & $-29$&$-32$ &\n&\n\\*
D& 0.70 & $-35$& 1228& 1134&   6.0&   9.5&   7   &  $-14$&
$-32$&$-139$&     0.8&   $ 9$\\*
C& 0.72 & $-57$&  683&   18&   9.9&  11.6&  $-8 $&  $-20$&  $-139$& $-97$&    1.1&   $12$\\*
B& 0.92 & $-69$ &  454&  290&  9.1&   9.9&    40&   $29$&  $-97$& $-45$&    1.0&    $6 $\\*
A& 2.41 & $-88$ &  557&  224&  4.1&   7.1&     4 &  $-7 $&
$-135$&$-135$ &   1.3&    $1$\\

\sidehead{2145+067}
C& 0.13 &   72 &   \n& 6486&  \n&   4.1&\n&  $10$&    71&123&\n&\n\\*
B& 0.63 &  113 &  709&  703&   6.9&   6.9&   $26$&  $10 $&   123&139& 0.9&   $7$\\*
A& 0.86 &  121 & 1662&  566&   6.8&   7.7&   $24$&  $-2 $&   139& 139&
0.9&   $ 9$\\
\sidehead{2201+315}
f& 0.18 & $-153$ &   \n&  119&  \n&\n&\n&\n& $-154$&$-154$&\n&\n\\*
E& 0.46 & $-154$ & 1397&  631&   2.4&   5.2&    83&  $-70$& $-154$& $-154$&    2.2&     10\\*
D& 0.86 & $-154$ &  484&  510&   6.4&   9.6&  $-18$&  $-21$& $-154$& $-82$&    1.2&    $6$\\*
C& 0.96 & $-141$ &  100&   36&   4.7&\n&  $-15$&\n&  $-82$& $-82$&    1.2&    $1$\\*
B& 3.28 & $-144$ &  134&   97&   7.0&\n&  $-32$&\n& $-134$& $-106$&    2.2& $2$\\*
A& 3.99 & $-137$ &  150&   50&\n&\n&\n&\n& $-106$&$-134$&\n&\n\\ 
 
\enddata
\tablenotetext{a}{Position derived from 22 GHz data.}
\tablecomments{Columns are as follows: (1) Component
name. (2) Distance from core at 43 GHz, in milliarcseconds. (3)
Position angle with respect to core at 43 GHz. (4--5) Flux density at
22 and 43 GHz [mJy]. (6--7) Percentage polarization at 22 and 43 GHz,
measured at position of I peak. (8-9) Electric vector position angle
at 22 and 43 GHz, measured at position of I peak. (10-11) Local jet
direction, measured up- and downstream of the component.  (12)
Ratio of 43 and 22 GHz fractional polarizations, calculated using
convolved 43 GHz image. (13) Difference in electric vector position
angle between 43 and 22 GHz, calculated
using convolved 43 GHz image. }

\end{deluxetable}

\begin{deluxetable}{llc}

\tablecolumns{3}
\tablecaption{\label{ks_results} High-Polarization Versus
Low-Polarization Radio Quasars}
\tablewidth{0pt}
\tablehead{ \colhead{Property}& \colhead{K-S
Probability\tablenotemark{a}} & \colhead{Significant?}}

\startdata
Redshift &  0.60 & N\\
Optical V magnitude & 0.96 & N \\
Spectral index between 1.6 and 5 GHz & 0.60 & N\\
 Spectral index between 22 and 43 GHz  &   0.007 & Y\\
Total luminosity at 22 GHz  & 0.41 & N\\
Total luminosity at 43 GHz  & 0.25 & N\\
 Core luminosity at 43 GHz  &  0.08 & Y\\
 Ratio of core to extended flux at 43 GHz &   0.02 & Y \\
Total amount of jet bending out to 15 pc & 0.60 & N\\
 Jet misalignment between parsec- and kiloparsec-scales & 0.01& Y\\
Parsec-scale integrated percentage polarization at 43 GHz & 0.60 & N\\
 Percentage polarization of core at 43 GHz &   0.003 & Y\\
Offset of optical $\chi$ from innermost jet direction  & 0.60 & N\\
Percentage polarization of jet components at 43 GHz & 0.43 & N\\
Projected distance of jet components from the core & 0.80&N \\
 Luminosity of jet components at 43 GHz &  0.08 & Y\\
 Offset of component $\chi$'s from upstream jet direction &  0.0009 &Y\\
 Offset of component $\chi$'s from downstream jet direction &  0.00011&Y \\

\enddata
\tablenotetext{a}{Probability that HPQ and LPQ samples are  drawn from the same population,
according to a Kolmogorov-Smirnov test.}
\end{deluxetable}

\end{document}